\documentclass[11pt]{article}
\usepackage{amssymb}

\usepackage{graphicx}
\usepackage{amsmath}
\usepackage{makeidx}
\usepackage{indentfirst}

\newcounter{resultnum}[section]

\newcounter{conclusionnum}[section]

\newcounter{conditionnum}[section]

\newcounter{conjecturenum}[section]

\newcounter{examplenum}[section]

\newcounter{exercisenum}[section]

\newcounter{lemmanum}[section]

\newcounter{notationnum}[section]

\newcounter{theoremnum}[section]

\newcounter{definitionnum}[section]

\newcounter{corollarynum}[section]

\newcounter{remarknum}[section]

\newcounter{propositionnum}[section]

\newcounter{acknowledgementnum}[section]

\newcounter{algorithmnum}[section]

\newcounter{axiomnum}[section]

\newcounter{casenum}[section]

\newcounter{claimnum}[section]

\newcounter{summarynum}[section]

\newcounter{problemnum}[section]

\begin{document}

\title{Nonholomic Distributions and \\
Gauge Models of Einstein Gravity}
\date{September 26, 2009}
\author{ Sergiu I. Vacaru\thanks{
sergiu.vacaru@uaic.ro, Sergiu.Vacaru@gmail.com;\newline   http://www.scribd.com/people/view/1455460-sergiu }
\\
%EndAName
{\quad} \\
{\small {\textsl{ Science Department, University "Al. I. Cuza" Ia\c si},}
}\\
{\small {\textsl{\ 54 Lascar Catargi  street, 700107, Ia\c si, Romania}} }}
%%%%%%%%%%%%%%%%%%%%%%%%%
\maketitle

\begin{abstract}
For (2+2)--dimensional nonholonomic distributions, the physical information contained into a spacetime (pseudo) Riemannian metric can be encoded equivalently into new types of geometric structures and linear connections constructed as nonholonomic deformations of the Levi--Civita connection. Such deformations and induced geometric/physical objects are completely
determined by a prescribed metric tensor. Reformulation of the Einstein equations in nonholonomic variables (tetrads and new connections, for instance, with constant coefficient curvatures and/or Yang--Mills like potentials) reveals hidden geometric and rich quantum structures. It is shown how the Einstein gravity theory can be re--defined equivalently as certain gauge models on nonholonomic affine and/or de Sitter frame bundles.
We speculate on possible applications of the geometry of nonholonomic distributions with associated nonlinear connections in classical and quantum gravity.

\vskip0.3cm



\end{abstract}

\newpage
\tableofcontents

\section{Introduction}

It is a well known fact that, in a certain sense, the Einstein theory of
gravity, i.e. general relativity, can be viewed as a gauge model for the
non--semisimple Poincar\' e/affine, group and, in another sense, it can also
be considered as a gauge theory of the group of spacetime translations,
which are equivalent to arbitrary diffeomorphisms. Details of such and
alternative gauge gravity constructions, relevant to this paper, are
reviewed in \cite{pd1,pd2,tseytl,hehl,sard,ali}.\footnote{%
there were published some tenths of monographs and thousands of articles on
"gauge gravity models";\ we are not able to discuss here a number of
important ideas and results or to provide a comprehensive list of references}

In the standard approach to general relativity, a (pseudo) Riemannian metric
$\mathbf{g}$ and the corresponding Levi--Civita connection $\ ^{\mathbf{g}%
}\nabla $ are taken as basic variables\footnote{%
we use left ''up/low'' labels in order to emphasize that some
geometric/physical objects are defined by the same fundamental geometric
structures, see explanations in next section} representing the gravitational
field and its interactions. The Einstein equations were originally
formulated for geometric objects defined by such standard $\left( \mathbf{g,}%
\ ^{\mathbf{g}}\nabla \right) $ variables. Nevertheless, various purposes in
classical gravity (for instance, description of gravitational interactions
with spinor fields and computation of observable gravitational and light
interaction effects) and different attempts to quantize Einstein gravity and
formulate generalizations of gravity theory employed the tetrad/spinor
fields and different connection forms as (new) fundamental variables, see
discussions in Refs. \cite{asht,rov,thiem1}.

The dynamics of gravitational fields for a four dimensional spacetime is
defined by six independent components (from general ten ones) of a symmetric
metric field. Working with different variables in classical and quantum
models of Einstein gravity, we have to consider additional, in general,
nonholonomic (equivalently, nonintegrable and/or anholonomic) constraints on
fundamental field equations and elaborate geometric models in certain forms
involving nonholonomic structures. The mathematical formalism for physical
theories with constrained dynamics has its origins in the concept of
nonholonomic manifold and geometrization of nonholonomic mechanics \cite%
{vr1,hor,vr1,vr2}, see a review of results and historical details in \cite%
{bejf,vers}. Following methods of Finsler geometry and constructions adapted
to nonholonomic distributions and nonlinear connections, it was developed an
approach to geometrization of Lagrange and Hamilton mechanics and higher
order generalizations \cite{ma,mhh}.\footnote{%
There were elaborated various approaches to geometrization of nonholonomic
mechanics and classical and quantum field theories. We cited some monographs
related to metric compatible constructions, proposed by E. Cartan, and
developed by R. Miron's school on Lagrange--Hamilton and Finsler--Cartan
geometries, and genereralizations, following the nonlinear connection
formalism.}

In a series of works, we proved that nonholonomic variables and methods of
nonlinear connection geometry (here we note that nonholonomic
Lagrange--Finsler variables can be considered even in general relativity)
are very effective for developing new geometric methods for constructing
exact solutions in gravity theories, analysis of nonlinear solitonic
gravitational and matter field interactions, elaborating noncommutative and
gauge like generalizations etc, see reviews of results in \cite%
{ijgmmp,vrflg,vncg,vacap,ancv}.\footnote{%
We emphasize that in this article we shall not work with more general
classes of Lagrange--Finsler geometries elaborated in original form on
tangent bundles \cite{ma} but apply the nonholonomic manifold geometric
formalism in classical gravity and further developments to quantum gravity.}
Using nonholonomic distributions on (pseudo) Riemannian manifolds
(Lagrange--Finsler, or Hamilton--Cartan, ones and their almost K\"{a}hler
structures), we showed how the Einstein gravity can be quantized \cite%
{vpla,vfqlf,vqg4,anav,vloopdq,vgwgr} following methods of Fedosov
(deformation) quantization \cite{fed1,fed2,karabeg1}, loop quantum gravity %
\cite{asht,rov,thiem1} and/or brane A--model approach \cite{gukwit} (for our
purposes, such models where considered to enabled with nonholonmic
structures). Since the successful standard models in particle physics are
related to gauge theories with Yang--Mills type potentials, it is reasonable
to expect that there may be constructed a gauge like model of quantum
gravity. The perturbation theory of gauge fields is also more familiar for
phenomenological physicists. During almost fifty years, this prompted
various re--examinations of general relativity from the gauge theoretical
points of view and related geometric methods. All such approaches resulted
in a general conclusion that if a viable gauge like theory of gravity can be
elaborated, it will not be a usual Yang--Mills theory.

The bulk of models on gauge gravity were constructed as certain
generalizations, or alternatives, to the Einstein gravity theories.
Nevertheless, among thousands of works on gauge gravity, two papers by D.
Popov and L.\ Dikhin \cite{pd1,pd2} play a more special role. They proved
that the Yang--Mills equations for the so--called Cartan connection 1--form
(in the bundle of affine frames) projected on the base spacetime manifold
are just the Einstein equations for general relativity. Nevertheless, that
gauge gravity theory (and different similar models, for instance, for the
Poincar\'{e} gauge group) was considered 'unphysical' because the affine
structure group is non--semisimple, i.e. with degenerated Killing form,
which results in 'nonvariational' field equations in the total bundle.
Perhaps, that was not a substantial problem because the Yang--Mills
equations can be formally derived by ''pure'' geometric methods using a
formal dualization of 2--forms of curvature and/or by introducing an
auxiliary bilinear form on the fiber space. Projecting such gauge gravity
equations on the base, it is possible to obtain the same Einstein equations
not depending on any auxiliary fiber values. Together with the fact that
such constructions had not provided additional tools which could solve the
problem of renormalizability of gravitational interactions, that motivated a
number of researches to develop different types of generalized models of
gauge gravity with nonmeticity and/or dynamical torsion and extensions of
the affine structural group to the de Sitter and/or another ones, nonlinear
gauge symmetries, Higgs like broken symmetries of gauge group etc \cite%
{pd1,pd2,tseytl,hehl,sard,ali}. Various sophisticate mechanisms to extract,
or to get in certain limits, the classical Einstein gravity were proposed.

Our approach to gauge gravity is different from the former ones. We attempt
to construct a gauge like theory in certain bundle spaces enabled with
nonholonomic distributions (for simplicity, defining some classes of
nonholonomic frames with associated nonlinear connection structure) which
will encode equivalently all geometric and physical data for the Einstein
equations on (pseudo) Riemannian spaces. So, we are not going to generalize
the Einstein theory but try to reformulate it in certain nonholonomic
variables lifted, for instance on affine, or de Sitter bundles. Lifts of
geometric/physical objects will be performed in such forms that a new
geometric techniques will be possible to be applied in order to develop a
formal renormalization scheme of gravitational interactions, see second
partner work \cite{vpqgg}. We shall use the nonlinear connection formalism
and anholonomic frame method which we elaborated for certain gauge models
with generalizations of the Einstein gravity to Lagrange--Finsler
structures, higher order bundles and noncommutative gravity \cite%
{vgon,vd,vncg1,vncg}. Nevertheless, we emphasize that in this work the
nonholonomic geometric techniques will be considered only for bundle spaces
on (pseudo) Riemannian manifolds, in particular, on Einstein spacetimes.

\vskip2pt Our purpose is twofold:

\begin{enumerate}
\item To prove that the Einstein theory of gravity can be alternatively
formulated in terms of nonholonomic deformations of linear and nonlinear
connection structures uniquely\footnote{%
but equivalently to the well known approaches with Levi--Civita, Ashtekar
and various gauge like gravitational connections} defined by a fundamental
metric structure. The main issue will be related to constructions defining
nonholonomic frames and deformations of connections which result in constant
coefficient curvatures encoding the information for Einstein spaces.

\item To show that having introduced a corresponding nonholonomic
distribution the geometric and physical constructions with the ''standard''
Levi--Civita connection split into two categories of classical objects (and
quantum objects, see a partner paper \cite{vpqgg}): The first ones are in
terms of distinguished linear connections, which allows us to encode a part
of gravitational data into terms of some constant matrix curvatures, like in %
\cite{vacap}, or (equivalently) in terms of almost K\"{a}hler/
Lagrange--Finsler and/or other nonholonomic variables for (pseudo)
Riemannian manifolds \cite{vrflg,vpla,vqg4,vloopdq,vgwgr}. The second ones
contain contributions of a 'distortion' tensor which can be encoded into a
formal nonholonomic gauge gravity model (we shall analyze two such examples).%
\footnote{%
One should be emphasized that in this work any distinguished connection and
distorsion tensor will be completely defined by a metric structure,
similarly to the Levi--Civita connection. As a matter of principle, all
constructions with distinguished geometric objects can be equivalently
reformulated for standard variables in general relativity.}
\end{enumerate}

Let us now briefly describe the content of the present work. We outline the
geometry of nonholonomic distributions on (pseudo) Riemannian manifolds
(Sec. 2); the geometric and physical objects are adapted to an associated
nonlinear connection nonholonomic spacetime splitting when various classes
of metric compatible connections are defined by a metric structure. It is
shown how the Einstein gravity theory can be reformulated equivalently in
nonholonomic variables for different linear connection structures (Sec. 3).
We construct two models of gauge gravity in nonholonomic affine/ de Sitter
frame bundles, with Yang--Mills like equations which can be equivalently
re--defined into the standard Einstein equations for the general relativity
theory on the base spacetime (Sec. 4). We close with some concluding remarks
(Sec. 5). In Appendix, we present some necessary local formulas.

\section{Nonholonomic Distributions on (Pseudo) Riemannian Manifolds}

In this section we outline some new geometric features of general relativity
in order to fix the notations and provide a clear insight of nonholonomic
variables. Our goal is to prove that the Levi--Civita connection can be
decomposed conventionally into an auxiliary distinguished connection
(d--connection) structure and a distorsion tensor.\footnote{%
The complete system of field equations and constraints for an auxiliary
d--connection and distorsion tensor fields will be constructed to be
equivalent to the standard Einstein equations.} Such decompositions can be
performed in a canonical form for some nonholonomic distributions defining $%
2+2$ spacetime decompositions.

\subsection{Nonlinear connections and metrics}

We consider a four dimensional (in general, nonholonomic) pseudo--Riemannian
manifold (spacetime) $\mathbf{V,}$ endowed with a metric, $\mathbf{g},$ for
instance, of signature $(-,+,+,+),$ and the corresponding Levi--Civita
connection, $\ ^{\mathbf{g}}\nabla =\{\ _{\shortmid }^{\ ^{\mathbf{g}%
}}\Gamma _{\ \beta \gamma }^{\alpha }\},$ structures such that $\ ^{\mathbf{g%
}}\nabla \mathbf{g}=0$ and torsion $\ _{\shortmid }\mathcal{T}=\{\
_{\shortmid }T_{\ \beta \gamma }^{\alpha }\}$ of $\ ^{\mathbf{g}}\nabla $
vanishes, $\ _{\shortmid }\mathcal{T}=0.$\footnote{%
Definitions for different types of connections and torsions are given below,
in this section. Here we also note that we follow our system of notations,
see details in Refs. \cite{vrflg,ijgmmp,vgwgr}, when left ''up/low'' labels
show that (for instance) a value $\ ^{\mathbf{g}}\nabla =\nabla \lbrack \
\mathbf{g}]$ is completely defined by $\mathbf{g}.$ Any right indices are
usual abstract, or coordinate, tensor ones. In this work, a pair $(\mathbf{V}%
,\mathcal{N}),$ where $\mathcal{N}$ is a nonintegrable distribution on a
(pseudo) Riemannian spacetime $\mathbf{V}$ of enough smooth class, is called
a nonholonomic manifold.}

Local coordinates on $\mathbf{V}$ are denoted in the form $u^{\alpha
}=(x^{i},y^{a})$ (or, in brief, $u=(x,y)$) where indices of type $%
i,j,...=1,2 $ will be considered as formal horizontal/ holonomic ones
(h--indices), labeling h--coordinates, and indices of type $a,b,...=3,4$
will be considered as formal vertical/nonholonomic ones (v--indices),
labeling v--coordinates. We shall also use 'underlined' indices ($\underline{%
\alpha }=(\underline{i},\underline{a}),\underline{\beta }=(\underline{i},%
\underline{b}),...$), for local coordinate bases $e_{\underline{\alpha }%
}=\partial _{\underline{\alpha }}=(\partial _{\underline{i}},\partial _{%
\underline{a}}),$ equivalently $\partial /\partial u^{\underline{\alpha }%
}=(\partial /\partial x^{\underline{i}},\partial /\partial y^{\underline{a}%
});$ for dual coordinate bases we shall write \ $e^{\underline{\alpha }}=du^{%
\underline{\alpha }}=(e^{\underline{i}}=dx^{\underline{i}},e^{\underline{a}%
}=dx^{\underline{a}}).$ There will be considered primed indices ($\alpha
^{\prime }=(i^{\prime },a^{\prime }),\beta ^{\prime }=(j^{\prime },b^{\prime
}),...$), with double primes etc, for other local abstract/coordinate bases,
for instance, $e_{\alpha ^{\prime }}=(e_{i^{\prime }},e_{a^{\prime
}}),e^{\alpha ^{\prime }}=(e^{i^{\prime }},e^{a^{\prime }})$ and $e_{\alpha
^{\prime \prime }}=(e_{i^{\prime \prime }},e_{a^{\prime \prime }}),e^{\alpha
^{\prime \prime }}=(e^{i^{\prime \prime }},e^{a^{\prime \prime }}),$ where $%
i^{\prime },i^{\prime \prime },=1,2...$ and $a^{\prime },a^{\prime \prime
}=3,4.$

On a manifold $\mathbf{V,}$ we can introduce any nonholonomic distribution
\begin{equation*}
\mathbf{N}=N_{i}^{a}(u)dx^{i}\otimes \partial _{a}
\end{equation*}%
defined by a set of coefficients $N_{i}^{a}(u)=N_{i}^{a}(x,y).$\footnote{%
this is similar to the possibility to consider on a spacetime any frame
and/or coordinate systems, or any 2+2 and 3+1 decompositions} If such
coefficients state also a decomposition of the tangent bundle $T\mathbf{V}$
into conventional horizontal (h), $h\mathbf{V,}$ and vertical (v), $v\mathbf{%
V,}$ subspaces as a Whitney sum
\begin{equation}
T\mathbf{V}=h\mathbf{V}\oplus v\mathbf{V,}  \label{whit}
\end{equation}%
we say that $\mathbf{N}=\{N_{i}^{a}\}$ defines a nonlinear connection
(N--connection) on $\mathbf{V}$ and that such a nonholonomic (equivalently,
anholonomic) space is a N--anholonomic manifold.

To a N--connection structure, we can associate a class of vielbein (frame)
transforms which are linear on coefficients $N_{i}^{a},$
\begin{equation}
\mathbf{e}_{\alpha }^{\ \underline{\alpha }}=\left[
\begin{array}{cc}
e_{i}^{\ \underline{i}}(u) & N_{i}^{b}(u)e_{b}^{\ \underline{a}}(u) \\
0 & e_{a}^{\ \underline{a}}(u)%
\end{array}%
\right] ,\ \mathbf{e}_{\ \underline{\beta }}^{\beta }=\left[
\begin{array}{cc}
e_{\ \underline{i}}^{i\ }(u) & -N_{k}^{b}(u)e_{\ \underline{i}}^{k\ }(u) \\
0 & e_{\ \underline{a}}^{a\ }(u)%
\end{array}%
\right] ,  \label{vt}
\end{equation}%
where, in a particular case, $e_{i}^{\ \underline{i}}=\delta _{i}^{%
\underline{i}}$ and $e_{a}^{\ \underline{a}}=\delta _{a}^{\underline{a}},$
with $\delta _{i}^{\underline{i}}$ and $\delta _{a}^{\underline{a}}$ being
the Kronecker symbols. Such transform define the so--called N--adapted frame
and co--frame structures, respectively,$\mathbf{e}_{\alpha }=\mathbf{e}%
_{\alpha }^{\ \underline{\alpha }}\partial _{\underline{\alpha }}$ and $%
\mathbf{e}_{\ }^{\beta }=\mathbf{e}_{\ \underline{\beta }}^{\beta }du^{%
\underline{\beta }},$ when the N--elongated partial derivatives/ frames are
\begin{eqnarray}
\mathbf{e}_{\alpha } &\doteqdot &\delta _{\alpha }=\left( \delta
_{i},\partial _{a}\right)  \label{dder} \\
&\equiv &\frac{\delta }{\delta u^{\alpha }}=\left( \mathbf{e}_{i}=\frac{%
\delta }{\delta x^{i}}=\partial _{i}-N_{i}^{a}\left( u\right) \partial _{a},%
\frac{\partial }{\partial y^{a}}\right)  \notag
\end{eqnarray}%
and the N--elongated differentials / co--frames,
\begin{eqnarray}
\mathbf{e}_{\ }^{\beta } &\doteqdot &\delta \ ^{\beta }=\left( dx^{i},\delta
y^{a}\right)  \label{ddif} \\
&\equiv &\delta u^{\alpha }=\left( \delta x^{i}=dx^{i},\mathbf{e}_{\
}^{a}=\delta y^{a}=dy^{a}+N_{i}^{a}\left( u\right) dx^{i}\right) .  \notag
\end{eqnarray}%
There are used both type of denotations $\mathbf{e}_{\alpha }\doteqdot
\delta _{\alpha }$ and $\mathbf{e}_{\ }^{\beta }\doteqdot \delta \ ^{\beta }$
in order to preserve a relation to denotations from Refs. \cite%
{vrflg,vncg,ma}.\footnote{%
We adopt the convention that for the spaces provided with N--connection
structure the geometrical objects are denoted by ''boldfaced'' symbols if it
is necessary to distinguish such objects from similar ones for spaces
without N--connection.}

For (\ref{dder}), there are satisfied the anholonomy relations
\begin{equation}
\left[ \mathbf{e}_{\alpha },\mathbf{e}_{\beta }\right] =\mathbf{e}_{\alpha }%
\mathbf{e}_{\beta }-\mathbf{e}_{\beta }\mathbf{e}_{\alpha }=\mathbf{w}_{\
\alpha \beta }^{\gamma }\left( u\right) \mathbf{e}_{\gamma }  \label{anhr}
\end{equation}%
with nontrivial anholonomy coefficients $\mathbf{w}_{\beta \gamma }^{\alpha
}\left( u\right) $ computed as
\begin{equation}
\mathbf{w}_{~ji}^{a}=-\mathbf{w}_{~ij}^{a}=\Omega _{ij}^{a},\ \mathbf{w}%
_{~ia}^{b}=-\mathbf{w}_{~ai}^{b}=\partial _{a}N_{i}^{b},  \label{anhc}
\end{equation}%
where $\Omega _{ij}^{a}$ define the N--connection curvature coefficients%
\begin{equation}
\Omega _{ij}^{a}\doteqdot \mathbf{e}_{j}\left( N_{i}^{a}\right) -\mathbf{e}%
_{i}\left( N_{j}^{a}\right) .  \label{ncurv}
\end{equation}

On a N--anholonomic manifold $\mathbf{V,}$ we can re--define all geometric
objects in a form adapted to a splitting (\ref{whit}), with respect to bases
(\ref{dder}) and (\ref{ddif}) and theirs tensor products. In such cases,
there are used the terms distinguished objects (in brief, d--objects) and,
in particular cases, there are considered d--tensors, d--connections,
d--metrics etc.

Any metric structure $\mathbf{g}$ on $\mathbf{V}$ can be parametrized in
local coordinate form,
\begin{equation*}
\mathbf{g}=g_{\underline{\alpha }\underline{\beta }}du^{\underline{\alpha }%
}\otimes du^{\underline{\beta }}
\end{equation*}%
with
\begin{equation}
\underline{g}_{\alpha \beta }=g_{\underline{\alpha }\underline{\beta }}=%
\left[
\begin{array}{cc}
g_{ij}+N_{i}^{a}N_{j}^{b}h_{ab} & N_{j}^{e}h_{ae} \\
N_{i}^{e}h_{be} & h_{ab}%
\end{array}%
\right] ,  \label{ansatz}
\end{equation}%
or, equivalently, in N--adapted form, as a d--metric,

\begin{equation}
\mathbf{g}=\mathbf{g}_{\alpha \beta }\left( u\right) \mathbf{e}^{\alpha
}\otimes \mathbf{e}^{\beta }=g_{ij}\left( u\right) e^{i}\otimes
e^{j}+h_{ab}\left( u\right) \mathbf{e}^{a}\otimes \mathbf{e}^{b}.
\label{block2}
\end{equation}%
A metric, for instance, parametrized in the form (\ref{ansatz})\ is generic
off--diagonal if it can not be diagonalized by any coordinate transforms.
The anholonomy coefficients (\ref{anhc}) do not vanish for the a generic
off--diagonal metric (\ref{ansatz}) (equivalently, d--metric (\ref{block2})).

\subsection{The set of metric compatible d--connections}

A distinguished connection (in brief, d--connection) on a spacetime $\mathbf{%
V}$,
\begin{equation}
\mathbf{D}=(\ ^{h}D;\ ^{v}D)=\{\mathbf{\Gamma }_{\beta \gamma }^{\alpha
}=(L_{jk}^{i},L_{bk}^{a};C_{jc}^{i},C_{bc}^{a})\},  \label{dcon}
\end{equation}%
is a linear connection preserving under parallel transports the distribution
(\ref{whit}). In explicit form, the coefficients\ $\mathbf{\Gamma }_{\beta
\gamma }^{\alpha }$ are computed with respect to a N--adapted basis (\ref%
{dder}) and (\ref{ddif}) following formulas
\begin{equation}
\mathbf{\Gamma }_{\ \alpha \beta }^{\gamma }\left( u\right) =\left( \mathbf{D%
}_{\alpha }\delta _{\beta }\right) \rfloor \delta ^{\gamma }.  \label{dcon1}
\end{equation}%
The operations of h- and v-covariant derivations, $\
^{h}D_{k}=\{L_{jk}^{i},L_{bk\;}^{a}\}$ and $\
^{v}D_{c}=\{C_{jk}^{i},C_{bc}^{a}\}$ from (\ref{dcon}) are introduced as
corresponding h- and v--parametrizations of (\ref{dcon1}),\ $%
L_{jk}^{i}=\left( \mathbf{D}_{k}\delta _{j}\right) \rfloor d^{i},$\ $%
L_{bk}^{a}=\left( \mathbf{D}_{k}\partial _{b}\right) \rfloor \delta ^{a},$\ $%
C_{jc}^{i}=\left( \mathbf{D}_{c}\delta _{j}\right) \rfloor$ $d^{i},$\ $%
C_{bc}^{a}=\left( \mathbf{D}_{c}\partial _{b}\right) \rfloor \delta ^{a},$\
where ''$\rfloor "$ denotes the interior product defined by a metric $%
\mathbf{g}=\{\mathbf{g}_{\alpha \beta }\}$ and it inverse $\mathbf{g}^{-1}=\{%
\mathbf{g}^{\alpha \beta }\}.$

We can introduce the d--torsion and d--curvature of a d--connection 1--form $%
\mathbf{\Gamma }_{\ \alpha }^{\gamma }=\mathbf{\Gamma }_{\ \alpha \beta
}^{\gamma }\mathbf{e}^{\beta },$ respectively, following formulas
\begin{equation}
\ \mathcal{T}^{\alpha }\doteqdot \mathbf{De}^{\alpha }=\delta \mathbf{e}%
^{\alpha }+\mathbf{\Gamma }_{\ \ \beta }^{\gamma }\wedge \mathbf{e}^{\beta }
\label{dt}
\end{equation}%
and
\begin{equation}
\ \mathcal{R}_{\ \ \beta }^{\alpha }\doteqdot \mathbf{D\Gamma }_{\ \beta
}^{\alpha }=\delta \mathbf{\Gamma }_{\ \ \beta }^{\alpha }-\mathbf{\Gamma }%
_{\ \ \beta }^{\gamma }\wedge \mathbf{\Gamma }_{\ \ \gamma }^{\alpha },
\label{dc}
\end{equation}%
were "$\wedge$" is the anti--symmetric product of differential forms. The
component formulas for $\mathcal{T}^{\alpha }$ and $\ \mathcal{R}_{\ \ \beta
}^{\alpha }$ are presented in Appendix, see formulas (\ref{dtors}) and (\ref%
{dcurv}) and details in Refs. \cite{vrflg,ijgmmp,vncg,ma}.

By straightforward computations, one proves two important results (see \cite%
{vnslms} and references therein, for N--anholonomic manifolds with symmetric
and nonsymmetric metrics, and \cite{ma}, for vector bundles):

\begin{enumerate}
\item \textit{Kawaguchi's metrization:} To any fixed d--connection $\
_{\circ }\mathbf{D}$ we can associate a d--connection $\mathbf{D}$ being
compatible with a metric $\mathbf{g},$ when it is satisfied the condition $%
\mathbf{D}_{\mathbf{X}}\mathbf{g=0},$ for any d--vector $\mathbf{X=}X^{i}%
\mathbf{e}_{i}+X^{a}e_{a}$ $\in T\mathbf{V}.$ The coefficients for
d--connections are related by formulas%
\begin{eqnarray*}
L_{\ jk}^{i} &=&\ _{\circ }L_{\ jk}^{i}+\frac{1}{2}g^{im}\ _{\circ
}D_{k}g_{mj},\ L_{\ bk}^{a}=\ _{\circ }L_{\ bk}^{a}+\frac{1}{2}g^{ac}\
_{\circ }D_{k}g_{cb}, \\
C_{\ jc}^{i} &=&\ _{\circ }C_{\ jc}^{i}+\frac{1}{2}g^{im}\ _{\circ
}D_{c}g_{mj},\ C_{\ bc}^{a}=\ _{\circ }L_{\ bc}^{a}+\frac{1}{2}g^{ae}\
_{\circ }D_{c}g_{eb}.
\end{eqnarray*}

\item \textit{Miron's procedure: \ }The set of d--connections $\{\mathbf{D}%
\} $ satisfying the conditions $\mathbf{D}_{\mathbf{X}}\mathbf{g=0}$ for a
given $\mathbf{g}$ is defined by formulas%
\begin{eqnarray}
L_{\ jk}^{i} &=&\widehat{L}_{jk}^{i}+\ ^{-}O_{km}^{ei}\mathbf{Y}%
_{ej}^{m\,},\ L_{\ bk}^{a}=\widehat{L}_{bk}^{a}+\ ^{-}O_{bd}^{ca}\mathbf{Y}%
_{ck}^{d\,},  \label{mcdc} \\
C_{\ jc}^{i} &=&\widehat{C}_{jc}^{i}+\ ^{+}O_{jk}^{mi}\mathbf{Y}%
_{mc}^{k\,},\ C_{\ bc}^{a}=\widehat{C}_{bc}^{a}+\ ^{+}O_{bd}^{ea}\mathbf{Y}%
_{ec}^{d\,},  \notag \\
\mbox{ where } \ ^{\pm }O_{jk}^{ih}&=&\frac{1}{2}(\delta _{j}^{i}\delta
_{k}^{h}\pm g_{jk}g^{ih}),\ ^{\pm }O_{bd}^{ca}=\frac{1}{2}(\delta
_{b}^{c}\delta _{d}^{a}\pm g_{bd}g^{ca})  \label{obop}
\end{eqnarray}%
are the so--called the Obata operators; $\mathbf{Y}_{ej}^{m\,},\mathbf{Y}%
_{mc}^{k\,},\mathbf{Y}_{ck}^{d\,}$ and $\mathbf{Y}_{ec}^{d\,}$ are arbitrary
d--tensor fields and $\widehat{\mathbf{\Gamma }}_{\ \alpha \beta }^{\gamma
}=\left( \widehat{L}_{jk}^{i},\widehat{L}_{bk}^{a},\widehat{C}_{jc}^{i},%
\widehat{C}_{bc}^{a}\right) ,$ with
\begin{eqnarray}
\widehat{L}_{jk}^{i} &=&\frac{1}{2}g^{ir}\left( \mathbf{e}_{k}g_{jr}+\mathbf{%
e}_{j}g_{kr}-\mathbf{e}_{r}g_{jk}\right) ,  \label{candcon} \\
\widehat{L}_{bk}^{a} &=&e_{b}(N_{k}^{a})+\frac{1}{2}g^{ac}\left( \mathbf{e}%
_{k}g_{bc}-g_{dc}\ e_{b}N_{k}^{d}-g_{db}\ e_{c}N_{k}^{d}\right) ,  \notag \\
\widehat{C}_{jc}^{i} &=&\frac{1}{2}g^{ik}e_{c}g_{jk},\ \widehat{C}_{bc}^{a}=%
\frac{1}{2}g^{ad}\left( e_{c}g_{bd}+e_{c}g_{cd}-e_{d}g_{bc}\right)  \notag
\end{eqnarray}%
is the canonical d--connection uniquely defined by the coefficients of
d--metric $\mathbf{g=}[g_{ij},g_{ab}]$ and N--connection $\mathbf{N}%
=\{N_{i}^{a}\}$ in order to satisfy the conditions $\widehat{\mathbf{D}}_{%
\mathbf{X}}\mathbf{g=0}$ and $\widehat{T}_{\ jk}^{i}=0$ and $\widehat{T}_{\
bc}^{a}=0$ but with general nonzero values for $\widehat{T}_{\ ja}^{i},%
\widehat{T}_{\ ji}^{a}$ and $\widehat{T}_{\ bi}^{a},$ see formulas (\ref%
{dtors}) computed for the case $\mathbf{\Gamma }_{\ \alpha \beta }^{\gamma }=%
\widehat{\mathbf{\Gamma }}_{\ \alpha \beta }^{\gamma }.$
\end{enumerate}

d--Tensors $\mathbf{Y}_{ej}^{m\,},\mathbf{Y}_{mc}^{k\,},\mathbf{Y}%
_{ck}^{d\,} $ and $\mathbf{Y}_{ec}^{d\,}$ parametrize the set of metric
compatible d--connections, with a metric $\mathbf{g,}$ on a N--anholonomic
manifold $\mathbf{V}.$ Having prescribed any values of such d--tensors (we
can follow any geometric, or physical theoretical/experimental, arguments;
for instance, we can take some zero, or non--zero, constants), we get a
metric compatible d--connection $\ ^{\mathbf{g}}\mathbf{D}$ (\ref{mcdc})
completely defined by a (pseudo) Riemannian metric $\mathbf{g}$ (\ref{ansatz}%
) (equivalently, d--metric (\ref{block2})) because formulas (\ref{candcon})
and (\ref{obop}) depend only on coefficients of a d--metric and
N--connection structures. Here we note that even the N--connection
coefficients are contained in the coefficient form of formulas for
d--connections the provided constructions hold true for any $2+2$ splitting
of a spacetime $\mathbf{V}.$ Such constructions are coordinate free because
there are considered nonholonomic distributions which do not depend on the
type of local frame/coordinate systems and their transforms.

As a matter of principle, any (pseudo) Riemannian geometry can be
alternatively described in terms of a canonical d--connection \ $\ ^{\mathbf{%
g}}\widehat{\mathbf{D}}$\ (\ref{candcon}) and/or any d--connection$\ ^{%
\mathbf{g}}\mathbf{D=}\ ^{\mathbf{g}}\widehat{\mathbf{D}}+\ ^{\mathbf{g}}%
\mathbf{Z}$ if the distorsion d--tensor $\ ^{\mathbf{g}}\mathbf{Z}$ is
completely defined by a metric $\mathbf{g.}$\footnote{%
We can use any linear connection $^{\mathbf{g}}D=\ ^{\mathbf{g}}\widehat{%
\mathbf{D}}+\ ^{\mathbf{g}}Z,$ which, in general, may be not adapted to the
N--connection splitting (\ref{whit}) but subjected to the condition that a
distorsion tensor $\ ^{\mathbf{g}}Z$ is completely defined by metric
structure.} In a metric compatible and N--adapted case, $\ ^{\mathbf{g}}%
\mathbf{Z}$ is defined by torsion $\ \ ^{\mathbf{g}}\mathcal{T}^{\alpha }$
of $\ ^{\mathbf{g}}\mathbf{D}$ but this torsion is completely different from
the case of the Riemann--Cartan, string or gauge gravity theories with
additional field equations for an additional 'physical' torsion field. On
N--anholonomic (pseudo) Riemannian manifolds, a canonical d--torsion $\ ^{%
\mathbf{g}}\widehat{\mathcal{T}}^{\alpha }$ (in more general cases, any
d--torsion $\ ^{\mathbf{g}}\mathcal{T}^{\alpha }$) is generated as a
nonholonomic deformation effect, by anholonomic coefficients (\ref{dder}),
including curvature of N--connection (\ref{ncurv}), completely defined by
certain generic off--diagonal coefficients with $N_{i}^{a}$ in a metric $%
\mathbf{g.}$

Of course, all above presented geometric constructions can be equivalently
redefined in terms of (standard, for (pseudo) Riemannian geometry)
Levi--Civita connection $\ ^{\mathbf{g}}\nabla .$ Nevertheless, for various
purposes in deformation quantization, brane A--model quantization and
nonholonomic loop quantization of gravity theories and solitonic hierarchies
in Einstein and Lagrange--Finsler spaces, it is more convenient to work with
nonholonomic variables and generalized connections (we considered, for
instance, certain examples of $\ ^{\mathbf{g}}\mathbf{D}$ for almost K\"{a}%
hler variables or d--connections with constant curvature d--connections),
see details and discussions, respectively, in Refs. \cite%
{vpla,vfqlf,vqg4,vloopdq,anav} and \cite{vacap,ancv}.

\subsection{D--connections with constant coefficient curvatures}

Considering nonholonomic frames (vierbeins and, equivalently, tetrads) on $%
\mathbf{V}$ of type $e^{\alpha }=e_{\ \underline{\alpha }}^{\alpha }du^{%
\underline{\alpha }}$ and $e^{\alpha }=e_{\ \alpha ^{\prime }}^{\alpha }%
\mathbf{e}^{\alpha ^{\prime }},$ and their duals defined respectively by
matrices $e_{\alpha \ }^{\ \underline{\alpha }}$ and $e_{\alpha \ }^{\
\alpha ^{\prime }},$ we may represent any d--metric $\mathbf{g}$ and some
related classes of d--connections to be parametrized by constant
coefficients with respect to certain N--adapted bases. We should use
''boldface'' values, for instance, $\mathbf{e}_{\alpha \ }^{\ \underline{%
\alpha }}$ and $\mathbf{e}_{\alpha \ }^{\ \alpha ^{\prime }},$ if such
transforms deform smoothly, nonholonomically, a N--connection 2+2 splitting
into another similar one. For two d--metrics
\begin{equation}
\mathbf{g=g}_{\alpha \beta }\mathbf{e}^{\alpha }\otimes \mathbf{e}^{\beta }%
\mbox{ \ and \ }\mathbf{\mathring{g}=\mathring{g}}_{\alpha ^{\prime }\beta
^{\prime }}\mathbf{\mathring{e}}^{\alpha ^{\prime }}\otimes \mathbf{%
\mathring{e}}^{\beta ^{\prime }},  \label{freq}
\end{equation}%
we have $\mathbf{g=\mathring{g}}$ if and only if
\begin{equation}
\mathbf{g}_{\alpha \beta }\mathbf{e}_{\ \alpha ^{\prime }}^{\alpha }\mathbf{e%
}_{\ \beta ^{\prime }}^{\beta }=\mathbf{\mathring{g}}_{\alpha ^{\prime
}\beta ^{\prime }},  \label{freq1}
\end{equation}%
for some 'vierbeins' $\mathbf{e}_{\ \alpha ^{\prime }}^{\alpha }$ which can
be defined algebraically for any given values $\mathbf{g}_{\alpha \beta }$
and prescribed constant coefficients $\mathbf{\mathring{g}}_{\alpha ^{\prime
}\beta ^{\prime }},$ with respect to bases of type (\ref{ddif}) elongated,
respectively, by some $N_{i}^{a}(u)$ and $\mathring{N}_{i^{\prime
}}^{a^{\prime }}(u^{\prime })$ \ for $u^{\alpha ^{\prime }}=u^{\alpha
^{\prime }}(u^{\alpha }).$ In N--adapted form, we can parametrize $\mathbf{e}%
_{\ \alpha ^{\prime }}^{\alpha }=(e_{\ i^{\prime }}^{i},e_{\ a^{\prime
}}^{a})$ and write (\ref{freq}) and (\ref{freq1}) as%
\begin{eqnarray}
\mathbf{g} &=&g_{ij}e^{i}\otimes e^{j}+h_{ab}\mathbf{e}^{a}\otimes \mathbf{e}%
^{b}=\mathring{g}_{i^{\prime }j^{\prime }}e^{i^{\prime }}\otimes
e^{j^{\prime }}+\mathring{h}_{a^{\prime }b^{\prime }}\mathbf{\mathring{e}}%
^{a^{\prime }}\otimes \mathbf{\mathring{e}}^{b^{\prime }},  \notag \\
\mathbf{\mathring{e}}^{\alpha ^{\prime }} &=&(e^{i^{\prime }}=dx^{i^{\prime
}},\mathbf{\mathring{e}}^{a^{\prime }}=dy^{a^{\prime }}+\mathring{N}%
_{i^{\prime }}^{a^{\prime }}dx^{i^{\prime }}),  \notag \\
\mathbf{e}^{\alpha } &=&(e^{i}=dx^{i},\mathbf{e}^{a}=dy^{a}+N_{i}^{a}dx^{i}),
\notag \\
\mbox{with\ }g_{ij}e_{\ i^{\prime }}^{i}e_{\ j^{\prime }}^{j} &=&\mathring{g}%
_{i^{\prime }j^{\prime }},\ h_{ab}e_{\ a^{\prime }}^{a}e_{\ b^{\prime }}^{b}=%
\mathring{h}_{a^{\prime }b^{\prime }},\ N_{i}^{a}e_{\ i^{\prime }}^{i}e_{a\
}^{\ a^{\prime }}=\mathring{N}_{i^{\prime }}^{a^{\prime }},  \label{rediff}
\end{eqnarray}%
for $e^{i}=e_{\ i^{\prime }}^{i}\mathring{e}^{i^{\prime }}$ and $\mathbf{e}%
^{a}=e_{\ a^{\prime }}^{a}\mathbf{\mathring{e}}^{a^{\prime }}$ when it is
convenient to take $[\mathring{e}^{i^{\prime }},\mathring{e}^{j^{\prime
}}]=0.$

Introducing the coefficients $\mathbf{\mathring{g}}_{\alpha ^{\prime }\beta
^{\prime }}$ and $\mathring{N}_{i^{\prime }}^{a^{\prime }}$ (equivalently, $%
\mathbf{g}_{\alpha \beta }$ and $N_{i}^{a})$ into formulas (\ref{candcon}),
we can compute the N--adapted coefficients of canonical d--connection $%
\widehat{\mathbf{\mathring{\Gamma}}}_{\ \alpha ^{\prime }\beta ^{\prime
}}^{\gamma ^{\prime }}$ (equivalently, $\widehat{\mathbf{\Gamma }}_{\ \alpha
\beta }^{\gamma }).$ Here we note that having defined the frame transform
coefficients $\mathbf{e}_{\ \alpha ^{\prime }}^{\alpha }$ from (\ref{freq1}%
), we can consider the corresponding transformation law for any
d--connection 1--form:
\begin{equation}
\mathbf{\mathring{\Gamma}}_{~\beta ^{\prime }}^{\alpha ^{\prime
}}\rightarrow \mathbf{\Gamma }_{~\beta }^{\alpha }=\mathbf{e}_{\ \alpha
^{\prime }}^{\alpha }\mathbf{\mathring{\Gamma}}_{~\beta ^{\prime }}^{\alpha
^{\prime }}\mathbf{e}_{\beta \ }^{\ \beta ^{\prime }}+\mathbf{e}_{~\gamma
^{\prime }}^{\alpha }\mathbf{\mathring{e}}^{\prime }(\mathbf{e}_{\beta
}^{~\gamma ^{\prime }}),  \label{dctr}
\end{equation}%
where $\mathbf{\mathring{e}}^{\prime }=$ $\delta u^{\mu ^{\prime }}\mathbf{%
\mathring{e}}_{\mu ^{\prime }}.$

By straightforward computations (see details on proof of Proposition 2.1 in
Ref. \cite{vacap}), using formulas (\ref{candcon}) for $\widehat{\mathbf{%
\mathring{\Gamma}}}_{\ \alpha ^{\prime }\beta ^{\prime }}^{\gamma ^{\prime
}} $ we find that any (pseudo) Riemannian metric $\mathbf{g=\mathring{g}}$
on $\mathbf{V}$ defines a set of metric compatible d--connections with
constant coefficients of type
\begin{equation}
\ \widehat{\mathbf{\mathring{\Gamma}}}_{\ \alpha ^{\prime }\beta ^{\prime
}}^{\gamma ^{\prime }}=\left( \widehat{\mathring{L}}_{\ j^{\prime }k^{\prime
}}^{i^{\prime }}=0,\widehat{\mathring{L}}_{\ b^{\prime }k^{\prime
}}^{a^{\prime }}=const,\widehat{\mathring{C}}_{j^{\prime }c^{\prime
}}^{i^{\prime }}=0,\widehat{\mathring{C}}_{b^{\prime }c^{\prime
}}^{a^{\prime }}=0\right)  \label{ccandcon}
\end{equation}%
with respect to N--adapted frames (\ref{dder}) and (\ref{ddif}) for\ any $%
\mathbf{\mathring{N}}=\{\mathring{N}_{i^{\prime }}^{a^{\prime }}(x^{\prime
},y^{\prime })\}$ being a nontrivial solution of the system of equations%
\begin{equation}
2\ \widehat{\mathring{L}}_{b^{\prime }k^{\prime }}^{a^{\prime }}=\frac{%
\partial \mathring{N}_{k^{\prime }}^{a^{\prime }}}{\partial y^{b^{\prime }}}%
-\ \mathring{h}^{a^{\prime }c^{\prime }}\ \mathring{h}_{d^{\prime }b^{\prime
}}\frac{\partial \mathring{N}_{k^{\prime }}^{d^{\prime }}}{\partial
y^{c^{\prime }}}  \label{auxf1}
\end{equation}%
for any nondegenerate constant--coefficients symmetric matrix $\mathring{h}%
_{d^{\prime }b^{\prime }}$ and its inverse $\ \mathring{h}^{a^{\prime
}c^{\prime }}.$

Putting the coefficients $\widehat{\mathbf{\mathring{\Gamma}}}_{\ \alpha
^{\prime }\beta ^{\prime }}^{\gamma ^{\prime }}$ (\ref{ccandcon}) into
formulas (\ref{dcurv}), we obtain constant curvature coefficients if the
conditions (\ref{auxf1}) are satisfied:
\begin{eqnarray}
\ \widehat{\mathbf{\mathring{R}}}_{\ \beta ^{\prime }\gamma ^{\prime }\delta
^{\prime }}^{\alpha ^{\prime }} &=&(\widehat{\mathring{R}}_{~h^{\prime
}j^{\prime }k^{\prime }}^{i^{\prime }}=0,\widehat{\mathring{R}}_{~b^{\prime
}j^{\prime }k^{\prime }}^{a^{\prime }}=\ \widehat{\mathring{L}}_{\ b^{\prime
}j^{\prime }}^{c^{\prime }}\ \widehat{\mathring{L}}_{\ c^{\prime }k^{\prime
}}^{a^{\prime }}-\ \widehat{\mathring{L}}_{\ b^{\prime }k^{\prime
}}^{c^{\prime }}\ \widehat{\mathring{L}}_{\ c^{\prime }j^{\prime
}}^{a^{\prime }}=const,  \notag \\
&&\ \widehat{\mathring{P}}_{~h^{\prime }j^{\prime }a^{\prime }}^{i^{\prime
}}=0,\ \widehat{\mathring{P}}_{~b^{\prime }j^{\prime }a^{\prime
}}^{c^{\prime }}=0,\ \widehat{\mathring{S}}_{~j^{\prime }b^{\prime
}c^{\prime }}^{i^{\prime }}=0,\ \widehat{\mathring{S}}_{~b^{\prime
}d^{\prime }c^{\prime }}^{a^{\prime }}=0).  \label{ccdcc}
\end{eqnarray}%
From formula (\ref{dctr}) for N--adapted frame transforms of d--connection,
one follows that $\widehat{\mathbf{\mathring{\Gamma}}}_{\ \alpha ^{\prime
}\beta ^{\prime }}^{\gamma ^{\prime }}$ $\rightarrow \widehat{\mathbf{\Gamma
}}_{\ \alpha \beta }^{\gamma }(u)$ and $\ \widehat{\mathbf{\mathring{R}}}_{\
\beta ^{\prime }\gamma ^{\prime }\delta ^{\prime }}^{\alpha ^{\prime
}}\rightarrow \ \widehat{\mathbf{R}}_{\ \beta \gamma \delta }^{\alpha }(u)$
resulting in non--constant coefficients for such geometric objects defined
with respect to bases $\mathbf{e}_{\alpha }$ and $\mathbf{e}^{\alpha }.$

Nevertheless, even the curvature d--tensor$\ \widehat{\mathbf{R}}_{\ \beta
\gamma \delta }^{\alpha }(u),$ in general, does not have constant
coefficients, the corresponding scalar curvature (\ref{sdccurv}) is constant
both for $\widehat{\mathbf{\mathring{\Gamma}}}_{\ \alpha ^{\prime }\beta
^{\prime }}^{\gamma ^{\prime }}$ \ and $\widehat{\mathbf{\Gamma }}_{\ \alpha
\beta }^{\gamma }$ if the conditions (\ref{auxf1}) were fixed with respect
to a basis $\mathbf{\mathring{e}}_{\mu ^{\prime }}$ and $\mathbf{\mathring{e}%
}^{\mu ^{\prime }}.$ Really, using formulas (\ref{dctr}) and (\ref{freq1})
and introducing values (\ref{ccandcon}) and (\ref{ccdcc}) into (\ref{dricci}%
), we get:
\begin{eqnarray*}
\overleftrightarrow{\mathbf{\mathring{R}}}&\doteqdot& \ \mathbf{\mathring{g}}%
^{\alpha ^{\prime }\beta ^{\prime }}\ \ \widehat{\mathbf{\mathring{R}}}%
_{\alpha ^{\prime }\beta ^{\prime }}=\mathring{g}^{i^{\prime }j^{\prime }}\
\widehat{\mathring{R}}_{i^{\prime }j^{\prime }}+\ \mathring{h}^{a^{\prime
}b^{\prime }}\widehat{\mathring{S}}_{a^{\prime }b^{\prime }}=\
\overrightarrow{\mathring{R}}+\ \overleftarrow{\mathring{S}}=const \\
\mbox{ and } \overleftrightarrow{\mathbf{R}}&\doteqdot&\ \mathbf{g}^{\alpha
\beta }\ \ \widehat{\mathbf{R}}_{\alpha \beta }=\ \overrightarrow{R}+\
\overleftarrow{S}=\ \mathbf{\mathring{g}}^{\alpha ^{\prime }\beta ^{\prime
}}\ \ \widehat{\mathbf{\mathring{R}}}_{\alpha ^{\prime }\beta ^{\prime }}=%
\overleftrightarrow{\mathbf{\mathring{R}}}=const.
\end{eqnarray*}%
We conclude that using corresponding nonholonomic frame transform, for any
metric $\mathbf{g=\mathring{g}}$ on $\mathbf{V,}$ we can always construct a
d--connection with constant scalar curvature and it is also possible to
chose such nonholonomic constraints and N--adapted bases when the curvature
and Ricci d--tensors have constant coefficients. In Ref. \cite{vpqgg}, we
consider such constraints also for other curvature invariants which is
important for elaborating a formal procedure of renormalization by using
d--connections of type $\widehat{\mathbf{\mathring{\Gamma}}}_{\ \alpha
^{\prime }\beta ^{\prime }}^{\gamma ^{\prime }}$ (the above considered
d--tensors and d--connections are completely defined by a metric $\mathbf{g}$
but contain certain freedom in choosing necessary types of nonholonomic
distributions/transforms/frames).

Finally, we emphasize that the above mentioned geometric constructions can
not be performed for the Levi--Civita connection corresponding to a metric $%
\mathbf{g=\mathring{g}}$ on $\mathbf{V,}$ see details in next section.

\section{Einstein Gravity in N--adapted Variables}

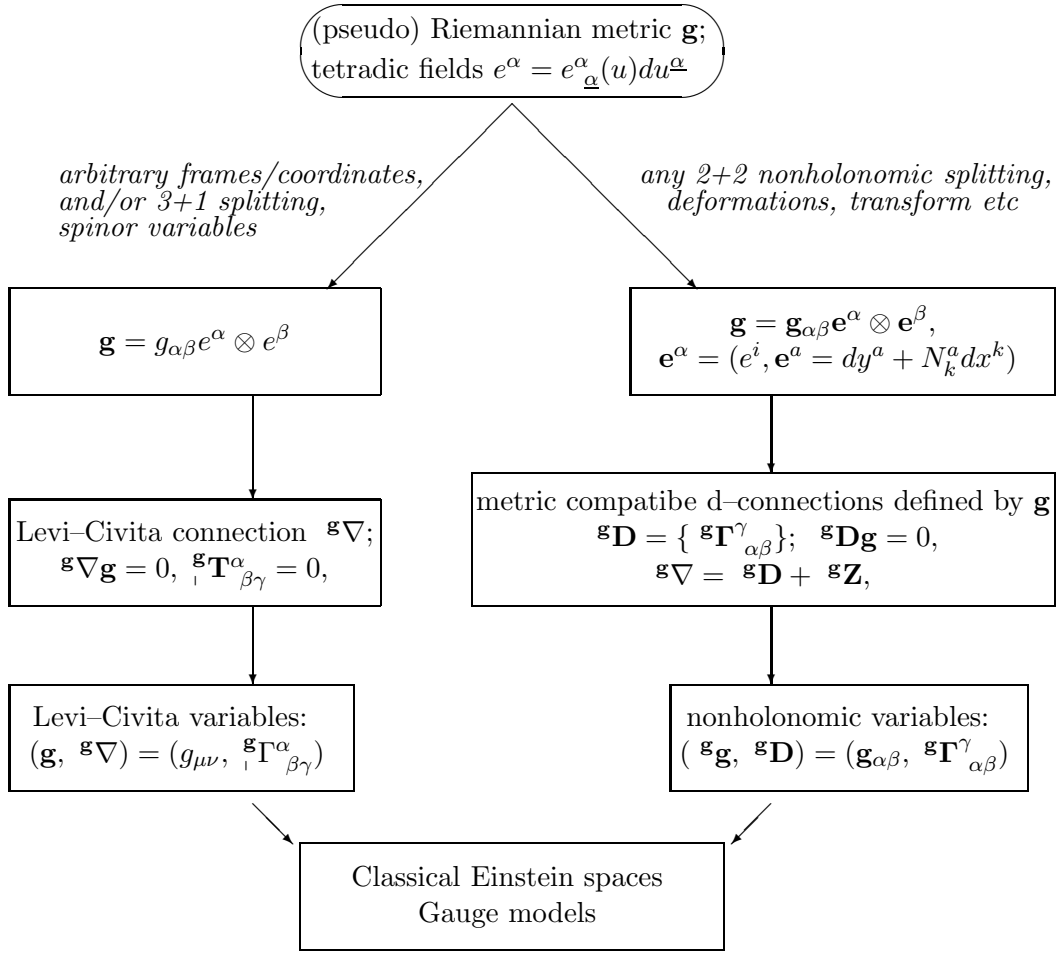
\begin{figure}[tbph]
\begin{center}
\begin{picture}(360,380)
\thinlines
\put(190,340){\oval(160,35)}
\put(114,345){\makebox{(pseudo) Riemannian metric ${\bf g};$}}

\put(114,330){\makebox{tetradic fields
$e^{\alpha }=e_{\ \underline{\alpha }}^{\alpha }(u) du^{\underline{\alpha }}$}}
\put(18,290){\makebox{\it arbitrary frames/coordinates,}}

\put(18,280){\makebox{\it and/or 3+1 splitting,}}

\put(18,270){\makebox{\it spinor variables}}
\put(238,290){\makebox{\it any 2+2 nonholonomic splitting,}}
\put(248,280){\makebox{\it deformations, transform etc}}
\put(0,210){\framebox(140,40){$\mathbf{g}=g_{\alpha \beta} e^\alpha \otimes e^\beta$}}
\put(235,210){\framebox(160,40)
{$\begin{array}{c}
  \mbox{$\mathbf{g}= \mathbf{g}_{\alpha \beta} \mathbf{e}^\alpha \otimes \mathbf{e}^\beta,$ }\\
 \mbox{$\mathbf{e}^\alpha = (e^i, \mathbf{e}^a = dy^a + N^a_k dx^k)$}
 \end{array}$
}}
\put(0,130){\framebox(140,40)
{$\begin{array}{c}
 \mbox{\ Levi--Civita connection $\ ^{\mathbf{g}}\nabla ;$}\\
 \mbox{$\ ^{\mathbf{g}}\nabla \mathbf{g}=0, \ _{\shortmid }^{\mathbf{g}}\mathbf{T}_{\ \beta \gamma }^{\alpha }=0,$}
 \end{array}$
 }}
\put(175,130){\framebox(220,50)
{$\begin{array}{c}
 \mbox{\ metric compatibe d--connections defined by $\mathbf{g}$}\\
 \mbox{$\
\ ^\mathbf{g}\mathbf{D}=\{\ ^\mathbf{g}\mathbf{\Gamma }_{\ \alpha \beta
}^{\gamma }\};\ \ ^\mathbf{g}\mathbf{D}\mathbf{g}=0,$}\\
 \mbox{$\ ^{%
\mathbf{g}}\nabla = \ ^\mathbf{g}\mathbf{D} +\ ^{\mathbf{g}}%
\mathbf{Z},$}
 \end{array}$
 }}
 \put(0,60){\framebox(130,40)
 {$\begin{array}{c}
 \mbox{Levi--Civita variables: }\\
 \mbox{$(\mathbf{g},\ ^{\mathbf{g}}\nabla )= ( g_{\mu \nu }, \ _{\shortmid }^{\mathbf{g}}%
\Gamma _{\ \beta \gamma }^{\alpha } )$}
 \end{array}$
 }}
   \put(250,60){\framebox(135,40)
{$\begin{array}{c}
 \mbox{nonholonomic variables: }\\
 \mbox{$(\ ^\mathbf{g}\mathbf{g} ,\ ^\mathbf{g}\mathbf{D}) = (\mathbf{g}_{\alpha \beta},\ ^\mathbf{g}\mathbf{\Gamma }_{\ \alpha \beta}^{\gamma } ) $}
 \end{array}$
 }}
\put(110,0){\framebox(160,40)
{$\begin{array}{c}
 \mbox{Classical Einstein spaces}\\
 \mbox{Gauge models}
 \end{array}$
}}

\put(288,210){\shortstack[r]{ \vector(0,-1){30}}}
\put(92,210){\shortstack[r]{ \vector(0,-1){40}}}
\put(288,130){\shortstack[r]{ \vector(0,-1){30}}}
\put(92,130){\shortstack[r]{ \vector(0,-1){30}}}
\put(288,55){\shortstack[r]{ \vector(-1,-1){15}}}
\put(92,55){\shortstack[r]{ \vector(1,-1){15}}}

\put(190,320){\shortstack[r]{\vector(1,-1){70}}}
\put(190,320){\shortstack[r]{\vector(-1,-1){70}}}

%\thicklines

\end{picture}
\end{center}
\caption{\textbf{Nonholonomic Variables in Gravity}}
\label{fig1}
\end{figure}

The 'standard' formulation of classical Einstein theory and first approaches
to quantum gravity were elaborated in variables $(\mathbf{g},\ ^{\mathbf{g}%
}\nabla ),$ where $\ ^{\mathbf{g}}\nabla =\nabla \lbrack \mathbf{g}]=\{\
_{\shortmid }^{\mathbf{g}}\Gamma _{\ \beta \gamma }^{\alpha }=\ _{\shortmid
}\Gamma _{\ \beta \gamma }^{\alpha }[\mathbf{g}]\}$ is the Levi--Civita
connection. This linear connection is completely constructed from the
coefficients of a metric $\mathbf{g=\{\mathbf{g}_{\mu \nu }\},}$ and their
first derivatives, on a spacetime manifold $\mathbf{V}$ following the
condition that $\ ^{\mathbf{g}}\nabla \mathbf{g}=0$ and $\ _{\shortmid }^{%
\mathbf{g}}T_{\ \beta \gamma }^{\alpha }=0,$ where $\ _{\shortmid }^{\mathbf{%
g}}T$ is the torsion of $\ ^{\mathbf{g}}\nabla .$\footnote{%
It should be noted that this connection is not adapted to the distribution (%
\ref{whit}) because it does not preserve under parallelism the h- and
v--distribution.} In another turn, for different purposes in classical and
quantum gravity, there were introduced various types of tetradic, or spinor,
variables and $3+1$ spacetime decompositions (for instance, in the
so--called Arnowit--Deser--Misner, ADM, formalism, Ashtekar variables and
loop quantum gravity), or nonholonomic $2+2$ splittings, see reviews of
results, discussion and references in \cite{rov,thiem1,asht}\ and \cite%
{vloopdq,vgwgr}, on nonholonmic variables, and Figure \ref{fig1}.

For a (pseudo) Riemannian metric $\mathbf{g,}$ we can construct an infinite
number of linear connections $\ ^{\mathbf{g}}D$ which are metric compatible,
$\ ^{\mathbf{g}}D\mathbf{g}=0,$ and completely defined by coefficients $%
\mathbf{g=\{\mathbf{g}_{\mu \nu }\}.}$ Of course, in general, the torsion $\
^{\mathbf{g}}T=\ _{D}T[\mathbf{g}]$ of a $\ ^{\mathbf{g}}D$ is not zero.%
\footnote{%
for a general linear connection, we do not use boldface symbols if such a
geometric object is not adapted to a prescribed nonholonomic distribution} \
Nevertheless, we can work equivalently both with $^{\mathbf{g}}\nabla $ and
any $\ ^{\mathbf{g}}D$ if the distorsion tensor $\ ^{\mathbf{g}}Z=Z[\mathbf{g%
}]$ \ from the corresponding connection deformation,
\begin{equation}
\ ^{\mathbf{g}}\nabla =\ ^{\mathbf{g}}D+\ ^{\mathbf{g}}Z,  \label{condeform}
\end{equation}%
is also completely defined by the metric structure $\mathbf{g}$ (in the
metric compatible cases, $\ ^{\mathbf{g}}Z$ is proportional to $\ ^{\mathbf{g%
}}T).$ So, any geometric construction/field equations etc performed in terms
of a connection $\ ^{\mathbf{g}}D$ can be rewritten in terms of $\ ^{\mathbf{%
g}}\nabla ,$ and inversely. It is a matter of convenience for some geometric
constructions or physical models to use one of the connections from an
infinite set of metric compatible ones $\{\ ^{\mathbf{g}}D\}$ and $\ ^{%
\mathbf{g}}\nabla .$

Let us consider two explicit examples:\ In previous section, we defined the
so--called canonical d--connection $\ ^{\mathbf{g}}\widehat{\mathbf{\Gamma }}%
_{\ \alpha \beta }^{\gamma }$ (\ref{ccandcon}). The formula (\ref{deflc})
from Appendix allows us to compute the coefficients of the Levi--Civita
connection $\ ^{\mathbf{g}}\nabla =\{\ _{\shortmid }^{\mathbf{g}}\Gamma _{\
\alpha \beta }^{\gamma }\}$ because the components of both $\ ^{\mathbf{g}}%
\widehat{\mathbf{\Gamma }}_{\ \alpha \beta }^{\gamma }$ and distorsion
d--tensor $\ _{\shortmid }^{\mathbf{g}}Z_{\ \alpha \beta }^{\gamma }$ (\ref%
{deft}) depend on coefficients of d--metric and N--connection (that why we
put the left up label $\mathbf{g}$ and use boldface symbols). Similar
splitting formulas hold true for the constant coefficient d--connection $%
\widehat{\mathbf{\mathring{\Gamma}}}_{\ \alpha ^{\prime }\beta ^{\prime
}}^{\gamma ^{\prime }}$ (\ref{ccandcon}) but we have to use $\mathbf{g=%
\mathring{g}}$ and $\mathbf{\mathring{N},}$ when the d--metric and
N--connection coefficients are recomputed following formulas (\ref{rediff})
and (\ref{freq1}).

It is possible to compute the distorsion d--tensor $\ ^{\mathbf{g}}\mathbf{Z}
$ for any metric compatible d--connection $\ ^{\mathbf{g}}\mathbf{D}$ for a
splitting of linear connections of type (\ref{condeform}), $\ ^{\mathbf{g}%
}\nabla =\ ^{\mathbf{g}}\mathbf{D}+\ ^{\mathbf{g}}\mathbf{Z}.$ This follows
from formulae (\ref{mcdc}) and (\ref{deflc}):%
\begin{equation}
\ ^{\mathbf{g}}\mathbf{Z=\ }_{\shortmid }^{\mathbf{g}}Z-\mathbf{OY,}
\label{distgf}
\end{equation}%
where $\mathbf{OY}$ denotes the terms proportional to Obata operators (\ref%
{obop}) and arbitrary d--tensor fields $\mathbf{Y=}\{Y_{ej}^{m\,},Y_{mc}^{k%
\,},Y_{ck}^{d\,},Y_{ec}^{d\,}\}.$ This emphasizes that in Riemann geometry a
specific 'gauge freedom' exists with respect to choosing a metric linear
connection being adapted to a 2+2 splitting. Following Miron's procedure,
such a freedom is parametrized by the set of d--tensor fields $\mathbf{Y.}$%
\footnote{%
As a matter of principle, we can also work with metric noncompatible
d--connections and then to use the above mentioned Kawaguchi's metrization,
which results in a similar 'gauge' like freedom for d--connections.
Nevertheless, to work directly with nonmetricity fields is not only a more
difficult technical problem but it is not clear how to provide rigorous
motivations following 'scenarios' of standard theories in physics, see more
discussions in Refs. \cite{ijgmmp,vrflg}.}

Let us first consider the usual Einstein equations defined in terms of a
Levi--Civita connection $\nabla $ on a spacetime $\mathbf{V}:$%
\begin{equation}
\ _{\shortmid }R_{\ \beta }^{\underline{\alpha }}-\frac{1}{2}(\ _{\shortmid }%
\overleftrightarrow{R}+\Lambda )\mathbf{e}_{\ \beta }^{\underline{\alpha }%
}=8\pi G\ ^{m}\mathbf{T}_{\ \beta }^{\underline{\alpha }},  \label{seeq}
\end{equation}%
where $\ _{\shortmid }R_{\ \beta }^{\underline{\alpha }}=\mathbf{e}_{\
\gamma }^{\underline{\alpha }}\ _{\shortmid }R_{\ \ \beta }^{\ \gamma },$ $\
_{\shortmid }\overleftrightarrow{R}$ is the scalar curvature of $\nabla ,$ $%
\mathbf{T}_{\ \beta }^{\underline{\alpha }}$ is the effective
energy--momentum tensor, $\Lambda $ is the cosmological constant, $G$ is the
Newton constant in the units when the light velocity $c=1,$ and the
coefficients $\mathbf{e}_{\ \beta }^{\underline{\alpha }}$ of tetradic
decomposition $\mathbf{e}_{\ \beta }=\mathbf{e}_{\ \beta }^{\underline{%
\alpha }}\partial /\partial u^{\underline{\alpha }}$ are defined by the
N--coefficients of the N--elongated operator of partial derivation, see (\ref%
{dder}).

Having chosen a metric compatible d--connection $\mathbf{D,}$ we compute the
Ricci d--tensor $\mathbf{R}_{\ \beta \gamma }$ and the scalar curvature $\
\overleftrightarrow{\mathbf{R}},$ see formulas (\ref{dricci}) and $\ $(\ref%
{sdccurv}) when $\widehat{\mathbf{D}}\mathbf{\rightarrow D.}$ For any such
d--connections, we can also postulate the (nonholonomic) filed equations
\begin{equation}
\mathbf{R}_{\ \beta }^{\underline{\alpha }}-\frac{1}{2}(\overleftrightarrow{%
\mathbf{R}}+\Lambda )\mathbf{e}_{\ \beta }^{\underline{\alpha }}=8\pi G%
\overleftrightarrow{\mathbf{T}}_{\ \beta }^{\underline{\alpha }},
\label{deinsteq}
\end{equation}%
but such equations (also defined by the metric structure) are not equivalent
to the Einstein equations in general relativity if the tensor $%
\overleftrightarrow{\mathbf{T}}_{\ \beta }^{\underline{\alpha }}$ does not
include contributions of distorsion d--tensor $\ ^{\mathbf{g}}\mathbf{Z}$ (%
\ref{distgf}) in a necessary form. We can use nonholonomic gravitational
equations (\ref{deinsteq}) in various types of generalized gravity theories
(like noncommutative and/or string gravity \cite{vncg}). Such equations,
with d--connections, and their solutions with additionally constrained
integral varieties happen to be also very useful in constructing new classes
of exact solutions in general relativity, parametrized by generic
off--diagonal matrices \cite{ijgmmp,vrflg}.

Introducing \ in (\ref{deinsteq}) the absolute antisymmetric tensor $%
\epsilon _{\alpha \beta \gamma \delta }$ and the effective source 3--form
\begin{equation*}
\mathcal{T}_{\ \beta }=\overleftrightarrow{\mathbf{T}}_{\ \beta }^{%
\underline{\alpha }}\ \epsilon _{\underline{\alpha }\underline{\beta }%
\underline{\gamma }\underline{\delta }}du^{\underline{\beta }}\wedge du^{%
\underline{\gamma }}\wedge du^{\underline{\delta }}
\end{equation*}%
and expressing the curvature tensor $\mathcal{R}_{\ \gamma }^{\tau }=\mathbf{%
R}_{\ \gamma \alpha \beta }^{\tau }\ \mathbf{e}^{\alpha }\wedge \ \mathbf{e}%
^{\beta }$ of $\ \mathbf{\Gamma }_{\ \beta \gamma }^{\alpha }=\ _{\shortmid
}\Gamma _{\ \beta \gamma }^{\alpha }-\ \mathbf{Z}_{\ \beta \gamma }^{\alpha
} $ as $\mathcal{R}_{\ \gamma }^{\tau }=\ _{\shortmid }\mathcal{R}_{\ \gamma
}^{\tau }-\mathcal{Z}_{\ \gamma }^{\tau },$ where $\ _{\shortmid }\mathcal{R}%
_{\ \gamma }^{\tau }$ $=\ _{\shortmid }R_{\ \gamma \alpha \beta }^{\tau }\
\mathbf{e}^{\alpha }\wedge \ \mathbf{e}^{\beta }$ is the curvature 2--form
of the Levi--Civita connection $\nabla $ and the distorsion of curvature
2--form $\mathcal{Z}_{\ \gamma }^{\tau }$ is defined by $\ \mathbf{Z}_{\
\beta \gamma }^{\alpha }$ (\ref{distgf}), we derive the equations (\ref%
{deinsteq}) (varying the action on components of $\mathbf{e}_{\ \beta },$
see details for $\mathbf{D=}\widehat{\mathbf{D}}$ in Ref. \cite{vloopdq}).
The gravitational field equations written in terms of an arbitrary metric
compatible d--connection $\mathbf{D}$ can be represented as 3--form
equations,%
\begin{eqnarray}
\epsilon _{\alpha \beta \gamma \tau } && \left( \mathbf{e}^{\alpha }\wedge
\mathcal{R}^{\beta \gamma }+\Lambda \mathbf{e}^{\alpha }\wedge \ \mathbf{e}%
^{\beta }\wedge \ \mathbf{e}^{\gamma }\right) =8\pi G\mathcal{T}_{\ \tau },
\label{einsteq} \\
\mbox{when\qquad }\mathcal{T}_{\ \tau } &=&\ ^{m}\mathcal{T}_{\ \tau }+\ ^{Z}%
\mathcal{T}_{\ \tau },  \notag \\
\ ^{m}\mathcal{T}_{\ \tau } &=&\ ^{m}\mathbf{T}_{\ \tau }^{\underline{\alpha
}}\epsilon _{\underline{\alpha }\underline{\beta }\underline{\gamma }%
\underline{\delta }}du^{\underline{\beta }}\wedge du^{\underline{\gamma }%
}\wedge du^{\underline{\delta }},  \notag \\
\ ^{Z}\mathcal{T}_{\ \tau } &=&-\left( 8\pi G\right) ^{-1}\mathcal{Z}_{\
\tau }^{\underline{\alpha }}\epsilon _{\underline{\alpha }\underline{\beta }%
\underline{\gamma }\underline{\delta }}du^{\underline{\beta }}\wedge du^{%
\underline{\gamma }}\wedge du^{\underline{\delta }},  \notag
\end{eqnarray}%
where $\ ^{m}\mathbf{T}_{\ \tau }^{\underline{\alpha }}$ is the matter
tensor field. It should be noted here that the equations (\ref{einsteq}) are
equivalent to the standard Einstein equations (\ref{seeq}) for the
Levi--Civita connection.

We conclude that all geometric and classical physical information in
Einstein gravity formulated for data 1] $(\mathbf{g,}\ \ ^{\mathbf{g}}\nabla
),$ can be transformed equivalently into geometric objects and constructions
for any nonholonomic variables 2] $(\mathbf{g,\ }\ ^{\mathbf{g}}\mathbf{D}),$
from an infinite set of metric compatible d--connections defined by the same
metric $\mathbf{g.}$ A formal scheme for general relativity sketching a
''dictionary'' between two equivalent geometric ''languages'' in general
relativity (with a Levi--Civita and a metric compatible d--connection ones)
is presented in Figure \ref{fig1}.

% Figure

\section{N--adapted Gauge Models of Einstein Gravity}

There are known gauge gravity models when an equivalent re--definition of
the Einstein equations in a Yang--Mills like form implies the
non--semisim\-pli\-ci\-ty of the gauge group. Such constructions were
performed for nonvariational theories on the total space of the bundle of
affine frames (to this class one belong the Poincar\'{e} and affine group
gauge gravity theories) \cite{pd1,pd2}. For more complete expositions
concerning generalizations of the Einstein gravity theory to gauge models
for the de Sitter gauge group and nonlinear realizations, with
metric--affine spaces, locally anisotropic structures, noncommutative
gravity etc, see, e.g., \cite{tseytl,hehl,sard,ali,vgon,vd,vncg1} and
references therein.

In our works \cite{vgon,vd,vncg1}, we followed a geometric calculus similar
to that developed in Refs. \cite{pd1,pd2,tseytl} but generalized for
different types of (higher order) locally anisotropic, supersymmetric, or
noncommutative generalizations of gravity. That allowed us to construct
certain classes of nonholonomic gauge gravity models, when projections of
the generalized Yang--Mills equations, on (pseudo) Riemannian spacetimes,
resulted in standard Einstein equations for general relativity and/or field
equations for (non) commutative/super Lagrange--Finsler models of gravity.

The goal of this section is to prove that using splitting of connections $\
^{\mathbf{g}}\nabla =\ ^{\mathbf{g}}\mathbf{D}+\ ^{\mathbf{g}}\mathbf{Z},$
see formulas (\ref{condeform}) and (\ref{distgf}), the Einstein gravity
theory can be encoded equivalently into some nonholonomic gauge
gravitational theories when the Yang--Mills equations are defined by
gravitational equations (\ref{einsteq}), but projections on a nonholonomic
base result in vacuum gravitational equations (\ref{deinsteq}). Such models
provide a new perspective for renormalization perturbative schemes for the
Einstein gravity following a two connections formalism (when both
connections are defined by the same metric structure with respect to a
prescribed nonholonomic distribution) \cite{vpqgg}.

\subsection{Affine gauge distributions induced by Einstein metrics}

In our approach, we follow a global geometric calculus elaborated in Ref. %
\cite{bishop}. For Yang--Mills gauge fields on bundles spaces and related
models of gauge gravity, the formalism was developed in Ref. \cite{pd1,pd2}.
In nonholonomic forms, for nonholonomic gauge models of Lagrange and Finsler
anisotropic gravity and higher order generalizations, on (higher order)
vector/tangent bundles enabled with N--connection structure, such
computations were performed both in global and component forms in Refs. \cite%
{vgon,vd}, see also reviews of results, on gauge d--groups and d--spinors,
in \cite{vrflg,vncg,vjhep,vstav}. We shall omit details on computations,
which on N--anholonomic manifolds are very similar to those on nonholonomic
vector/tangent bundles, presenting some most important results necessary for
our further considerations in quantum gravity \cite{vpqgg}.

\subsubsection{Bundle of N--adapted linear frames}

Let $\mathbf{e}_{\alpha }=(\mathbf{e}_{i},e_{a})$ (\ref{dder}) be a
N--adapted frame at a point $u=(x,y)\in \mathbf{V.}$\footnote{%
We note that coordinates are defined for local carts on an atlas covering $%
\mathbf{V}$ and bundle spaces on such N--anholonomic manifolds; for
simplicity, we shall omit labels pointing that certain geometric objects are
defined on an open region/local cart $\mathcal{U}\subset \mathbf{V}$ and
write, for instance, instead of $\ _{\mathcal{U}}\mathbf{e}_{\alpha }$ and $%
\ _{\mathcal{U}}\mathbf{N,}$ only $\mathbf{e}_{\alpha }$ and $\mathbf{N}$
etc.} We can adapt the frame transform to a 2+2 splitting defined by a
N--connection $\mathbf{N}=\{N_{i}^{a}\}$ if there are considered local right
distinguished actions $\mathbf{e}_{\alpha ^{\prime }}=\mathbf{e}_{\alpha
^{\prime }}^{\ \alpha }\mathbf{e}_{\alpha }$ with matrices of type%
\begin{equation*}
\mathbf{e}_{\alpha ^{\prime }}^{\ \alpha }(u)=\left(
\begin{array}{cc}
\mathbf{e}_{i^{\prime }}^{\ i}(u) & 0 \\
0 & \mathbf{e}_{a^{\prime }}^{\ a}(u)%
\end{array}%
\right) \subset GL_{2+2}=GL(2,\mathbb{R})\oplus GL(2,\mathbb{R}),
\end{equation*}%
when $\mathbf{e}_{i^{\prime }}=\mathbf{e}_{i^{\prime }}^{\ i}\mathbf{e}_{i}$
and $e_{a^{\prime }}=\mathbf{e}_{a^{\prime }}^{\ a}e_{a},$ for $\mathbf{e}%
_{\alpha ^{\prime }}=(\mathbf{e}_{i^{\prime }},e_{a^{\prime }}).$ We denote
by $\ ^{N}\mathcal{L}a(\mathbf{V})$ $= \left( \ ^{N}La(\mathbf{V}),
GL_{2+2},\mathbf{V}\right)$ the bundle of linear adapted frames on $\mathbf{%
V}$ defined as the principal bundle with a surjective map $\ ^{N}\pi :\
^{N}La(\mathbf{V})\rightarrow \mathbf{V}$ transforming any adapted frame $%
\mathbf{e}_{\alpha ^{\prime }}$ in a point $u$ on base spacetime. The
nonholonomic total space$\ ^{N}La(\mathbf{V})$ \ (with nonholonomic
structure induced by $\mathbf{N}$ on $\mathbf{V)}$ is constructed as the set
of adapted frames $\mathbf{e}_{\alpha ^{\prime }}$ in all points of base
N--anholonomic manifold $\mathbf{V}$ enabled with a d--metric structure (\ref%
{block2}).

The structural (linear distinguished) group $GL_{2+2}$ (in brief, d--group)
acts on a typical fiber space. We denote by $\varrho _{\check{\alpha}%
}=(\varrho _{\check{i}},\varrho _{\check{a}})$ a basis in the Lie algebra $%
\emph{gl}_{2+2}$ of d--group $GL_{2+2}$ (N--adapting, we get a structural
Lie d--algebra) satisfying the commutation relations of type
\begin{equation*}
\left[ \varrho _{\check{\alpha}},\varrho _{\check{\beta}}\right] =\mathbf{f}%
_{\ \check{\alpha}\check{\beta}}^{\check{\gamma}}\varrho _{\check{\gamma}},%
\mbox{\ disinguished as\  }\left[ \varrho _{\check{i}},\varrho _{\check{j}}%
\right] =f_{\ \check{i}\check{j}}^{\check{k}}\varrho _{\check{k}},\ \left[
\varrho _{\check{a}},\varrho _{\check{b}}\right] =f_{\ \check{a}\check{b}}^{%
\check{c}}\varrho _{\check{c}},
\end{equation*}%
with structural constants $\mathbf{f}_{\ \check{\alpha}\check{\beta}}^{%
\check{\gamma}}=\left( f_{\ \check{i}\check{j}}^{\check{k}},f_{\ \check{a}%
\check{b}}^{\check{c}}\right) ,$ when conventional h-- and v--splitting have
to be introduced in order to perform N--adapted geometric constructions.
\footnote{%
It is not obligatory to split the fiber and total space constructions even a
base is a N--anholonomic manifold. Nevertheless, to elaborate gauge like
models of Einstein gravity, we have to consider pull--backs of geometric
objects from the base spacetime manifold which positively results in
d--objects with distinguishing of Lie group/algebras indices.} \ The Killing
form (fiber metric) is defined
\begin{equation}
\mathbf{K}(\varrho _{\check{\alpha}},\varrho _{\check{\beta}})=\mathbf{K}_{%
\check{\alpha}\check{\beta}}=\mathbf{f}_{\ \check{\beta}\check{\mu}}^{\check{%
\gamma}}\mathbf{f}_{\ \check{\alpha}\check{\gamma}}^{\check{\mu}}.
\label{killing}
\end{equation}%
For semisimple Lie groups, this metric is not degenerate which allows us to
define d--metrics on corresponding principal bundles induced by a d--metric $%
\mathbf{g}\ $(\ref{block2}) and Killing forms (\ref{killing}), $\mathbf{k}=%
\mathbf{g+K.}$\footnote{%
Bundle $^{N}\mathcal{L}a(\mathbf{V})=\left( \ ^{N}La(\mathbf{V})\mathbf{,}%
GL_{2+2},\mathbf{V}\right) $ is associated to $\left( \mathbf{P},\ ^{N}\pi ,%
\mathbf{V}\right) .$} Using the d--metrics $\mathbf{g}$ on $\mathbf{V}$ and $%
\mathbf{k}$ on $\ ^{N}La(\mathbf{V})\mathbf{,}$ we can introduce two
operators $\ ^{\mathbf{g}}\ast $ and $\ ^{\mathbf{g}}\widehat{\delta },$
acting on the space of differential $r$--forms $\Lambda ^{r}(\mathbf{V})$ on
$\mathbf{V.}$ Such operators are corresponding horizontal projections on the
base manifold of the total space operators $\ ^{\mathbf{k}}\ast $ and $\ ^{%
\mathbf{k}}\widehat{\delta },$ acting on the space of differential forms $%
\Lambda ^{r}(\ ^{N}La(\mathbf{V}))$ on $\ ^{N}La(\mathbf{V})\mathbf{,}$ for $%
r=1,2,3, $ if $\dim \mathbf{V}=4.$

In a simple form, for instance, the operators acting on a base are defined
by their actions on an orthonormalized basis $\mathbf{e}_{\alpha ^{\prime
\prime }}=(\mathbf{e}_{i^{\prime \prime }},e_{a^{\prime \prime }}),$
constructed by a N--adapted transform $\mathbf{e}_{\alpha ^{\prime \prime
}}^{\ \alpha }=\left( \mathbf{e}_{i^{\prime \prime }}^{\ i},\mathbf{e}%
_{a^{\prime ^{\prime }}}^{\ a}\right) ,$ when $\mathbf{g}=\sum\limits_{\mu
^{\prime \prime }=1}^{4}\eta (\mu ^{\prime \prime })\mathbf{e}^{\mu ^{\prime
\prime }}\otimes \mathbf{e}^{\mu ^{\prime \prime }},$ where $\eta
(1)=-1,\eta (2)=1,\eta (3)=1,\eta (4)=1.$ For $\ ^{\mathbf{g}}\ast :\Lambda
^{r}(\mathbf{V})\rightarrow \Lambda ^{4-r}(\mathbf{V}),$ we have
\begin{eqnarray*}
^{\mathbf{g}}\ast : \left( \mathbf{e}^{\mu _{1}^{\prime \prime }}\wedge
\ldots \wedge \mathbf{e}^{\mu _{r}^{\prime \prime }}\right) &\rightarrow
&-sign\left(
\begin{array}{ccccccc}
1 & 2 & \ldots & r & r+1 & \ldots & 4 \\
\mu _{1}^{\prime \prime } & \mu _{2}^{\prime \prime } & \ldots & \mu
_{r}^{\prime \prime } & \nu _{1}^{\prime \prime } & \ldots & \nu
_{4-r}^{\prime \prime }%
\end{array}%
\right) \\
&&\times \mathbf{e}^{\nu _{1}^{\prime \prime }}\wedge \ldots \wedge \mathbf{e%
}^{\nu _{4-r}^{\prime \prime }}
\end{eqnarray*}%
and the adjoint to absolute differential operator $d$ (acting on
differential form), associated to the scalar product for forms, specified
for $r$--forms, is
\begin{equation*}
\ ^{\mathbf{g}}\widehat{\delta }=(-1)^{r}\ ^{\mathbf{g}}\ast ^{-1}\circ
d\circ \ ^{\mathbf{g}}\ast ,
\end{equation*}%
for $\ ^{\mathbf{g}}\ast ^{-1}=-(-1)^{r(4-r)}\ ^{\mathbf{g}}\ast $ and '$%
\circ $' denoting supperpositions of operators. Introducing corresponding
canonical orthonormalized N--adapted bases on $\ ^{N}La(\mathbf{V})\mathbf{,}
$ we can define in a similar form the actions of operators $\ ^{\mathbf{k}%
}\ast $ and $\ ^{\mathbf{k}}\widehat{\delta },$ see details in \cite%
{pd1,vgon}.

We induce a d--connection 1-form $\mathbf{\omega }$ on $\ ^{N}La(\mathbf{V})$
by a pull--back of a metric compatible d--connection $\mathbf{D}$ (\ref{dcon}%
) on $\mathbf{V,}$%
\begin{equation}
\mathbf{\omega }^{\check{\alpha}}=\{\mathbf{\omega }_{\qquad \gamma
}^{\alpha ^{\prime \prime }\beta ^{\prime \prime }}\doteqdot \mathbf{\Gamma }%
_{\ \beta ^{\prime \prime }\gamma }^{\alpha ^{\prime \prime }}=\mathbf{e}_{\
\alpha }^{\alpha ^{\prime \prime }}\mathbf{\Gamma }_{~\beta \gamma }^{\alpha
}\mathbf{e}_{\beta ^{\prime \prime }\ }^{\ \beta }+\mathbf{e}_{~\tau
}^{\alpha ^{\prime \prime }}\mathbf{e}_{\gamma }(\mathbf{e}_{\beta ^{\prime
\prime }}^{~\tau })\},  \label{onedf}
\end{equation}%
with a formal splitting of Lie d--algebra indices as $\check{\alpha}%
\rightarrow (\alpha ^{\prime \prime }\beta ^{\prime \prime }).$ If the
coefficients $\mathbf{\Gamma }_{\ \beta \gamma }^{\alpha }$ are related to a
nonholonomic deformation of a Levi--Civita connection $\ _{\shortmid }\Gamma
_{\ \beta \gamma }^{\alpha },$ following formula (\ref{condeform}), we get a
nonholonomic lift of $\ ^{\mathbf{g}}\nabla $ on $\mathbf{V}$ into a $\ ^{%
\mathbf{\omega }}\mathbf{D=\{\mathbf{\omega }^{\check{\alpha}}\}}$ on $\
^{N}La(\mathbf{V}).$ The curvature of d--connection $\mathbf{\omega }$ is
computed%
\begin{equation}
\ ^{\mathbf{\omega }}\mathcal{R}=d\mathbf{\omega +\omega \wedge \omega }=%
\mathbf{\varrho }_{\alpha ^{\prime \prime }\beta ^{\prime \prime }}\otimes \
^{\mathbf{\omega }}\mathcal{R}_{\qquad \mu \nu }^{\alpha ^{\prime \prime
}\beta ^{\prime \prime }}\mathbf{e}^{\mu }\wedge \mathbf{e}^{\nu },
\label{curvlfb}
\end{equation}%
where the coefficients$\ \ ^{\mathbf{\omega }}\mathcal{R}_{\qquad \mu \nu
}^{\alpha ^{\prime \prime }\beta ^{\prime \prime }}=\mathbf{R}_{\ \beta
^{\prime \prime }\mu \nu }^{\alpha ^{\prime \prime }}=\mathbf{e}_{~\alpha
}^{\alpha ^{\prime \prime }}\mathbf{e}_{\beta ^{\prime \prime }}^{~\beta }%
\mathbf{R}_{\ \beta \mu \nu }^{\alpha }$ are computed following formulas (%
\ref{dcurv}) and $\mathbf{\varrho }_{\alpha ^{\prime \prime }\beta ^{\prime
\prime }}=\left(
\begin{array}{cc}
\rho _{i^{\prime \prime }j^{\prime \prime }} & 0 \\
0 & \rho _{a^{\prime \prime }b^{\prime \prime }}%
\end{array}%
\right) $ being the standard basis for the Lie d--aldebra of matrices $\emph{%
gl}_{2+2},$ when $\left( \rho _{i^{\prime \prime }j^{\prime \prime }}\right)
_{k^{\prime \prime }l^{\prime \prime }}=\delta _{i^{\prime \prime }k^{\prime
\prime }}\delta _{j^{\prime \prime }l^{\prime \prime }}$ and $\left( \rho
_{a^{\prime \prime }b^{\prime \prime }}\right) _{c^{\prime \prime }d^{\prime
\prime }}=\delta _{a^{\prime \prime }c^{\prime \prime }}\delta _{b^{\prime
\prime }d^{\prime \prime }}.$ \ For the 1--form (\ref{onedf}), we can also
compute another 1--form on $\ ^{N}La(\mathbf{V}),$%
\begin{equation}
\bigtriangleup \ ^{\mathbf{\omega }}\mathcal{R}\mathbf{\equiv }\ ^{\mathbf{k}%
}\widehat{\delta }\ ^{\mathbf{\omega }}\mathcal{R}\mathbf{+}\ ^{\mathbf{k}%
}\ast ^{-1}\left[ \mathbf{\mathbf{\omega }},\ ^{\mathbf{k}}\ast \ ^{\mathbf{%
\omega }}\mathcal{R}\right] ,  \label{onecdf}
\end{equation}%
where $\left[ \mathbf{\mathbf{\omega }},\ ^{\mathbf{k}}\ast \mathbf{\omega }%
\right] =\mathbf{\mathbf{\omega }}\wedge \ ^{\mathbf{k}}\ast \mathbf{\omega -%
}\ ^{\mathbf{k}}\ast \mathbf{\omega \wedge \mathbf{\omega ,}}$ and verify
that $d\ ^{\mathbf{\omega }}\mathcal{R+}\left[ \mathbf{\mathbf{\omega }},\ ^{%
\mathbf{\omega }}\mathcal{R}\right] =0.$ One defines the operator $%
\bigtriangleup =\widehat{H}$ $\circ \ ^{\mathbf{k}}\widehat{\delta },$ where
$\widehat{H}$ is the operator of horizontal projection on base.\footnote{%
A form $\varpi $ on a principal bundle $P$ with values in a Lie algebra $%
\mathcal{G}$ is called horizontal if $\widehat{H}\varpi =\varpi .$} \
Locally, the 1--form (\ref{onecdf}) is computed%
\begin{equation*}
\bigtriangleup \ ^{\mathbf{\omega }}\mathcal{R}=\mathbf{\varrho }_{\alpha
^{\prime \prime }\tau ^{\prime \prime }}\otimes \mathbf{g}^{\mu \lambda
}\left( \mathbf{D}_{\lambda }\ ^{\mathbf{\omega }}\mathcal{R}_{\qquad \mu
\nu }^{\alpha ^{\prime \prime }\tau ^{\prime \prime }}+f_{\qquad \beta
^{\prime \prime }\delta ^{\prime \prime }\ \gamma ^{\prime \prime
}\varepsilon ^{\prime \prime }}^{\alpha ^{\prime \prime }\tau ^{\prime
\prime }}\mathbf{\omega }_{\qquad \lambda }^{\beta ^{\prime \prime }\delta
^{\prime \prime }}\ ^{\mathbf{\omega }}\mathcal{R}_{\qquad \mu \nu }^{\gamma
^{\prime \prime }\varepsilon ^{\prime \prime }}\right) \mathbf{e}^{\nu }.
\end{equation*}
We get that the 1--form $\bigtriangleup \ ^{\mathbf{\omega }}\mathcal{R}$
vanishes for a standard curvature of a gauge $GL$--field. The last two
equations can be written in abstract form
\begin{eqnarray}
^{\mathbf{\omega }}\mathbf{D\ ^{\mathbf{\omega }}\mathcal{R}} &\mathbf{=}&0,
\label{bianchid} \\
\bigtriangleup \ ^{\mathbf{\omega }}\mathcal{R} &=&0,  \label{yangm}
\end{eqnarray}%
where the structure equations (\ref{bianchid}) \ are just the Biachi
identities and the field equations \ (\ref{yangm}) are just the Yang--Mills
equations for the structure (gauge) d--group $GL_{2+2}.$\footnote{%
We derived the Yang--Mills equations following ''pure'' geometric methods;
for semisimple structure groups, an equivalent variational method also holds
true, see Refs. \cite{pd1,vgon}.}

If \ $^{\mathbf{\omega }}\mathcal{R}$ is defined by $\ ^{\mathbf{g}}\mathbf{R%
}_{\ \beta \mu \nu }^{\alpha }$ determined by a (pseudo) Riemannian metric
on $\mathbf{V,}$ we get a 1--form \ $\bigtriangleup \ ^{\mathbf{\omega }}%
\mathcal{R}$ on $\ ^{N}La(\mathbf{V})$ induced nonholonomically by the
d--metric structure on the base N--anholonomic manifold. $\ $ Nevertheless,
by lifts and nonholonomic deformations on the bundle of linear N--adapted
frames $\ ^{N}La(\mathbf{V}),$ we are still not able to generate certain
types of Yang--Mills equations which would be equivalent to the Einstein
equations on $\mathbf{V.}$ We have to extend the constructions to the bundle
of affine N--adapted frames.

\subsubsection{Bundle of N--adapted affine frames}

There is another natural (in our case, nonholonomic) bundle on a spacetime $%
\mathbf{V,}$ extending \ $^{N}\mathcal{L}a(\mathbf{V}),$ called the bundle
of N--adapted affine frames $\ ^{N}\mathcal{A}a(\mathbf{V})=\left( \ ^{N}Aa(%
\mathbf{V})\mathbf{,}Af_{2+2},\mathbf{V}\right) $ with the structure affine
d--group $Af_{2+2}=GL_{2+2}\otimes \mathbb{R}^{4}$ and the total space $\
^{N}Aa(\mathbf{V})$ consisting from the set of affine frames on base
spacetime $\mathbf{V}$ constructed such a way that $\ ^{N}\mathcal{A}a(%
\mathbf{V})$ is a principal bundle for $Af_{2+2}.$ The corresponding Lie
d--algebra of $Af_{2+2}$ is $af_{2+2}=\emph{gl}_{2+2}\oplus \mathbb{R}^{4}.$
So, any form $\overline{\Theta }$ on $^{N}\mathcal{A}a(\mathbf{V})$ can be
expressed as $\overline{\mathbf{\Theta }}=(\mathbf{\Theta },\ ^{\mathbf{e}}%
\mathbf{\Theta }),$ where $\mathbf{\Theta }$ is the $\emph{gl}_{2+2}$%
--component and $\ ^{\mathbf{e}}\mathbf{\Theta }$ is the $\mathbb{R}^{4}$%
--component.

Any d--connection 1--form $\mathbf{\omega }$ (\ref{onedf}) in \ $^{N}%
\mathcal{L}a(\mathbf{V})$ induces a (canonical) Cartan connection $\overline{%
\mathbf{\omega }}$ in $\ ^{N}\mathcal{A}a(\mathbf{V})$ which can be
represented as $i^{\ast }\overline{\mathbf{\omega }}=\left( \mathbf{\omega
,\ ^{\mathbf{e}}\chi }\right) ,$ where $i:\ ^{N}\mathcal{A}a\rightarrow \ $\
$^{N}\mathcal{L}a$ is the trivial reduction of bundles and the shifting form
$\mathbf{\ ^{\mathbf{e}}\chi =}e_{\hat{\alpha}}\otimes \chi _{\ \mu ^{\prime
\prime }}^{\hat{\alpha}}\mathbf{e}^{\mu ^{\prime \prime }},$ with $e_{\hat{%
\alpha}}$ being the standard basis in $\mathbb{R}^{4}$ and $\chi _{\ \mu
^{\prime \prime }}^{\hat{\alpha}}$ defining a frame decomposition of
d--metric $\mathbf{g}\ $(\ref{block2}), when $\mathbf{g}_{\alpha ^{\prime
\prime }\beta ^{\prime \prime }}=\chi _{\ \alpha ^{\prime \prime }}^{\hat{%
\alpha}}\chi _{\ \beta ^{\prime \prime }}^{\hat{\beta}}\eta _{\hat{\alpha}%
\hat{\beta}},$ for $\eta _{\hat{\alpha}\hat{\beta}}=diag[-1,1,1,1].$ For a
d--connection $\overline{\mathbf{\omega }},$ we can define the curvature
d--tensor in $\ ^{N}\mathcal{A}a,$%
\begin{equation}
\ ^{\mathbf{\omega }}\overline{\mathcal{R}}=d\overline{\mathbf{\omega }}%
\mathbf{+}\overline{\mathbf{\omega }}\mathbf{\wedge }\overline{\mathbf{%
\omega }},  \label{curvafb}
\end{equation}%
for which the $\emph{gl}_{2+2}$--, $\mathbb{R}^{4}$--components are written
in $\ ^{\mathbf{\omega }}\overline{\mathcal{R}}=(\ ^{\mathbf{\omega }}%
\mathcal{R},\ ^{\mathbf{\omega }}\mathcal{T}),$ where $\ ^{\mathbf{\omega }}%
\mathcal{R}$ is just the curvature (\ref{curvlfb}) and the torsion
\begin{equation*}
\ ^{\mathbf{\omega }}\mathcal{T}=d\mathbf{\ ^{\mathbf{e}}\chi +\omega \wedge
\ ^{\mathbf{e}}\chi -\ ^{\mathbf{e}}\chi \wedge \omega }=e_{\hat{\alpha}%
}\otimes \mathbf{T}_{\ \beta \gamma }^{\hat{\alpha}}\mathbf{e}^{\beta }%
\mathbf{\wedge e}^{\gamma }
\end{equation*}%
with $\mathbf{T}_{\ \beta \gamma }^{\hat{\alpha}}=\chi _{\ \mu ^{\prime
\prime }}^{\hat{\alpha}}\mathbf{e}_{\ \mu }^{\mu ^{\prime \prime }}\mathbf{T}%
_{\ \beta \gamma }^{\mu },$ when the coefficients $\mathbf{T}_{\ \beta
\gamma }^{\mu }$ are computed following formulas (\ref{dtors}). If $\mathbf{%
\omega }$ is induced by a metric compatible d--connection $\mathbf{D}$ (\ref%
{dcon}) on $\mathbf{V}$ related to a nonholonomic deformation of a
Levi--Civita connection $\ _{\shortmid }\Gamma _{\ \beta \gamma }^{\alpha },$
following formula (\ref{condeform}), the values $\overline{\mathbf{\omega }}$
and resulting $^{\mathbf{\omega }}\overline{\mathcal{R}}$ and $\ ^{\mathbf{%
\omega }}\mathcal{T}$ are induced by a d--metric $\mathbf{g}\ $(\ref{block2}%
).

By straightforward computations, using the total space formula (\ref{onecdf}%
) but for $\overline{\mathbf{\omega }}$ in $\ ^{N}\mathcal{A}a(\mathbf{V}),$
see similar details in Refs. \cite{pd1,vgon,vd} (those constructions are for
usual affine frame bundles and respective generalizations on higher order
Lagrange--Finsler spaces), we obtain
\begin{equation*}
\bigtriangleup \ ^{\mathbf{\omega }}\overline{\mathcal{R}}\mathcal{=}\left(
\bigtriangleup \ ^{\mathbf{\omega }}\mathcal{R},\mathcal{R}\tau +\mathcal{R}%
i\right) ,
\end{equation*}%
with the standard $\emph{gl}_{2+2}$--component $\bigtriangleup \ ^{\mathbf{%
\omega }}\mathcal{R}$ and the $\mathbb{R}^{4}$--component defined by the sum
of
\begin{equation*}
\mathcal{R}\tau =\ ^{\mathbf{g}}\widehat{\delta }\ ^{\mathbf{\omega }}%
\mathcal{T}+\ ^{\mathbf{g}}\ast ^{-1}\left[ \mathbf{\omega },\ ^{\mathbf{g}%
}\ast \ ^{\mathbf{\omega }}\mathcal{T}\right]
\end{equation*}%
and
\begin{equation*}
\mathcal{R}i=\ ^{\mathbf{g}}\ast ^{-1}\left[ \mathbf{\ ^{\mathbf{e}}\chi },\
^{\mathbf{g}}\ast \ ^{\mathbf{\omega }}\mathcal{R}\right] =-e_{\hat{\alpha}%
}\otimes \chi _{\ \alpha }^{\hat{\alpha}}\mathbf{g}^{\alpha \beta }\mathbf{R}%
_{\beta \nu }\mathbf{e}^{\nu },
\end{equation*}%
where the components of the Ricci d--tensor $\mathbf{R}_{\beta \nu }$ are
computed following formulas (\ref{dricci}).

The nonholonomic field equations (\ref{deinsteq}) for variables $(\mathbf{g,D%
}),$ where a d--connection $\mathbf{D}$ (\ref{dcon}) is metric compatible,
induce an equivalent system of nonholnoimic Yang--Mills equations for
N--adapted Cartan 1--form $\overline{\mathbf{\omega }}$ in $\ ^{N}\mathcal{A}%
a(\mathbf{V}),$%
\begin{equation}
\bigtriangleup \ ^{\mathbf{\omega }}\overline{\mathcal{R}}\mathcal{=}%
\overline{\mathcal{J}},  \label{afgym}
\end{equation}%
with the source $\overline{\mathcal{J}}=\left( \mathcal{J},\mathbf{\ ^{%
\mathbf{e}}}\mathcal{J}\right) .$ Taking $\mathbf{\ ^{\mathbf{e}}}\mathcal{%
J=J}\tau +\mathcal{J}i,$ for
\begin{equation}
\mathcal{J}i=-e_{\hat{\alpha}}\otimes \chi _{\ \alpha }^{\hat{\alpha}}%
\mathbf{g}^{\alpha \beta }\left( \overleftrightarrow{\mathbf{E}}_{\beta \nu
}-\frac{1}{2}\mathbf{g}_{\beta \nu }\overleftrightarrow{\mathbf{E}}\right)
\mathbf{e}^{\nu },  \label{afsourc}
\end{equation}%
where $\overleftrightarrow{\mathbf{E}}=$ $\overleftrightarrow{\mathbf{E}}_{\
\nu }^{\nu },$ for $\overleftrightarrow{\mathbf{E}}_{\beta \nu }=8\pi G%
\overleftrightarrow{\mathbf{T}}_{\ \beta }^{\underline{\alpha }}+\frac{%
\Lambda }{2}\mathbf{g}_{\beta \nu },$ with a spin source $\mathcal{S}_{\quad
\beta \mu }^{\alpha }$ for an arbitrary torsion $\mathcal{T}$ in $\mathcal{R}%
\tau $ and $\bigtriangleup \ ^{\mathbf{\omega }}\mathcal{R=J=}0,$ we
generate nonholonomic gauge gravitational equations induced by a
d--connection $\mathbf{D}.$ In local form, the equations (\ref{afgym}) are
usual gauge field equations for the 1--form
\begin{equation}
\overline{\mathbf{\omega }}=\ ^{\mathbf{g}}\overline{\Gamma }=\left(
\begin{array}{cc}
\ ^{\mathbf{g}}\mathbf{\Gamma }_{\quad \beta ^{\prime \prime }}^{\alpha
^{\prime \prime }} & l_{0}^{-1}\mathbf{e}^{\alpha ^{\prime \prime }} \\
0 & 0%
\end{array}%
\right) \mbox{\ and \ }\mathbf{\omega }=\{\ ^{\mathbf{g}}\mathbf{\Gamma }%
_{\quad \beta ^{\prime \prime }}^{\alpha ^{\prime \prime }}\},  \label{aymf}
\end{equation}%
for an arbitrary dimension constant $l_{0}^{-1},$ when instead of the
Killing form (\ref{killing}), which is degenerated for the affine structural
groups, $\mathbf{K}=0;$ we can consider any auxiliary bilinear form $\mathbf{%
a}_{\alpha ^{\prime \prime }\beta ^{\prime \prime }}$ in order to have a
well defined d--metric $\mathbf{k}=\mathbf{g+a}$ in the total space of $\
^{N}\mathcal{A}a(\mathbf{V}).$ Such a dependence on an auxiliary bilinear
form consists a particular case of constructions related to Miron's
procedure on N--adapted affine frame bundles when the pull--backs of
d--tensors $\mathbf{Y}_{ej}^{m\,},\mathbf{Y}_{mc}^{k\,},\mathbf{Y}%
_{ck}^{d\,} $ and $\mathbf{Y}_{ec}^{d\,},$ see (\ref{mcdc}), to the total
space are such way taken that the metric compatibility $\mathbf{D}_{\mathbf{X%
}}\mathbf{g=0}$ on $\mathbf{V} $ is generalized to a $^{\mathbf{\omega }}%
\overline{\mathbf{D}}_{\overline{\mathbf{X}}}\mathbf{k=0}$ for ant $%
\overline{\mathbf{X}}\in T\ ^{N}\mathcal{A}a(\mathbf{V}).$ This type of
constructions do not affect the physical/geometrical objects on physical
spacetime $\mathbf{V}$ because projections on a base spacetime do not depend
on components of $\mathbf{a.}$

Now, let us state the conditions for the source $\overline{\mathcal{J}}%
=\left( \mathcal{J},\mathbf{\ ^{\mathbf{e}}}\mathcal{J}\right) $ in (\ref%
{afgym}) when the nonholonomic Yang--Mills equations are equivalent to the
Einstein equations in nonholonomic variables (\ref{einsteq}). By a
straightforward computation using 1--form (\ref{afgym}), we get a source (%
\ref{afsourc}), when
\begin{eqnarray}
\overleftrightarrow{\mathbf{E}}_{\quad \mu }^{\alpha ^{\prime \prime }}
&\rightarrow &\tau _{\quad \mu }^{\alpha ^{\prime \prime }}=\widetilde{T}%
_{\quad \mu }^{\alpha ^{\prime \prime }}-\frac{1}{2}\delta _{\quad \mu
}^{\alpha ^{\prime \prime }}\widetilde{T},  \label{ident} \\
\widetilde{T}_{\quad \mu }^{\alpha ^{\prime \prime }} &=&8\pi G(\
^{m}T_{\quad \mu }^{\alpha ^{\prime \prime }}+\ ^{Z}T_{\quad \mu }^{\alpha
^{\prime \prime }})+\frac{\Lambda }{2}\mathbf{e}_{\quad \mu }^{\alpha
^{\prime \prime }},  \notag
\end{eqnarray}%
and $\mathcal{R}\tau $ is induced by d--torsion field $\ ^{\mathbf{g}}%
\mathbf{T}_{\ \beta \gamma }^{\mu }$ (\ref{dtors}) of a $\ \ ^{\mathbf{g}}%
\mathbf{D}=\ ^{\mathbf{g}}\nabla -\ ^{\mathbf{g}}\mathbf{Z,}$ subjected to
conditions (\ref{condeform}) and (\ref{distgf}). \ Such gauge equations,%
\begin{equation}
\bigtriangleup \ ^{\mathbf{\omega }}\overline{\mathcal{R}}\mathcal{=}\ ^{%
\mathbf{g}}\overline{\mathcal{J}},  \label{geinsteq}
\end{equation}%
with $\ ^{\mathbf{g}}\overline{\mathcal{J}}=\left( 0,\mathbf{\ ^{\mathbf{e}}}%
\mathcal{J}=\ \ ^{\mathbf{g}}\mathcal{J}\tau +\ ^{\mathbf{g}}\mathcal{J}%
i\right) ,$ where $\ ^{\mathbf{g}}\mathcal{J}i$ is defined by $\tau _{\quad
\mu }^{\alpha ^{\prime \prime }},\ \ ^{\mathbf{g}}\mathcal{J}\tau =\ ^{%
\mathbf{g}}\widehat{\delta }\ ^{\mathbf{g}}\mathcal{T}+\ ^{\mathbf{g}}\ast
^{-1}\left[ \mathbf{\omega },\ ^{\mathbf{g}}\ast \ ^{\mathbf{g}}\mathcal{T}%
\right] ,$ for $\ ^{\mathbf{g}}\mathcal{T}=e_{\hat{\alpha}}\otimes \ ^{%
\mathbf{g}}\mathbf{T}_{\ \beta \gamma }^{\hat{\alpha}}\mathbf{e}^{\beta }%
\mathbf{\wedge e}^{\gamma },$ are also equivalent to the standard Einstein
equations (\ref{seeq}). For trivial N--anholonomic structures, with
vanishing d--torsion (\ref{dtors}), nonholonomy coefficients (\ref{anhc})
and N--connection curvature (\ref{ncurv}), the equations (\ref{geinsteq})
transform into similar ones derived by Popov and Dikhin \cite{pd2}, for the
Levi--Civita connection $\ _{\shortmid }^{\mathbf{g}}\mathbf{\Gamma }_{\quad
\beta ^{\prime \prime }}^{\alpha ^{\prime \prime }},$ when $\ ^{\mathbf{g}}%
\mathbf{\Gamma }_{\quad \beta ^{\prime \prime }}^{\alpha ^{\prime \prime
}}\rightarrow $ $\ _{\shortmid }^{\mathbf{g}}\mathbf{\Gamma }_{\quad \beta
^{\prime \prime }}^{\alpha ^{\prime \prime }}$ in (\ref{afgym}). The
Einstein equations written in such a gauge like form are
\begin{equation}
\bigtriangleup \ _{\shortmid }^{\mathbf{\omega }}\overline{\mathcal{R}}%
\mathcal{=}\ _{\shortmid }^{\mathbf{g}}\overline{\mathcal{J}},  \label{pdeq}
\end{equation}%
where $\ _{\shortmid }^{\mathbf{g}}\overline{\mathcal{J}}$ is constructed
following formulas (\ref{ident}) for distorsion $\ ^{Z}T_{\quad \mu
}^{\alpha ^{\prime \prime }}=0$ and $\ _{\shortmid }^{\mathbf{g}}\mathcal{J}%
\tau =0$ because $\ _{\shortmid }^{\mathbf{g}}\mathbf{T}_{\ \beta \gamma }^{%
\hat{\alpha}}=0$ and $\ _{\shortmid }^{\mathbf{g}}\mathcal{R}\tau =0.$

The gravitational Yang--Mills equations (\ref{pdeq}) in the bundle of affine
fra\-mes $\ \mathcal{A}a(V)=\left( \ Aa(V)\mathbf{,}Af_{4},V\right) $ on a
(pseudo) Riemannian spacetime $V$ not enabled with a nonholonomic
distribution structure, can be derived using ''pure'' geometric methods
acting with operators ''$d"$ and $\ ^{g}\widehat{\delta }$ on the Cartan
1--form defined by the Levi--Civita connection on the base $V.$ On total
spaces, such equations are not variational because of degenerate Killing
form for the affine group. Nevertheless, they can be written in a formal
variational form by introducing an auxilliar bilinear symmetric form $%
a_{\alpha ^{\prime \prime }\beta ^{\prime \prime }}$ like we discussed
above; projecting on a base manifold one transforms the gauge equations into
standard Einstein equations.

In the language of nonholonomic distributions on (pseudo) Riemannian
manifolds and principal bundles (with various types of structure groups,
semisimple and non--semisimple ones) on such manifolds, geometrically
derived nonholonomic Yang--Mills equations in total spaces are subjected to
certain types of non--integrable constraints. It is not surprising that
violations of former prescribed gauge symmetries may result in some formal
nonvariational gauge theories (as a matter of principle, such theories also
can be quantized following almost standard methods, see Ref. \cite{lyakh}).
Nevertheless, because the equations (\ref{geinsteq}) are equivalent to (\ref%
{pdeq}), and both types of nonholonomic and holonomic gravitational
Yang--Mills equations are equivalent to the standard Einstein equations (\ref%
{seeq}), quantization of such gauge gravity models contains the same
renormalizability problems as for the standard perturbative quantum approach
to general relativity\footnote{%
see Ref. \cite{vpqgg} with a proposal how to perform a formal
renormalization scheme of Einstein gravity using nonholonomic distributions
and bi--connections defined by the same metric structures}.

\subsection{De Sitter nonholonomic Einstein distributions}

We can elaborate another type of variational gauge gravity model with an
extension of the structural group. Geometrically, such constructions are
quite similar to the nonholonmic 'nonvariational' constructions on $\ ^{N}%
\mathcal{A} a(\mathbf{V})$ $= \left( \ ^{N}Aa(\mathbf{V})\mathbf{,}Af_{2+2},%
\mathbf{V}\right) ,$ when the data $(\mathbf{g,}\ \ ^{\mathbf{g}}\nabla )$
and/or $(\mathbf{g,\ }\ ^{\mathbf{g}}\mathbf{D})$ for the Einstein equations
induces equivalently the data $\overline{\mathbf{\omega }}=\ ^{\mathbf{g}%
}\Gamma $ solve a Yang--Mills equation. In general, the so--called affine
and de Sitter gauge gravities (see, for instance, reviews \cite%
{tseytl,hehl,sard,ali,vncg1}) are not equivalent to the Einstein gravity.
One has to consider various types of mechanisms with nonlinear realizations,
to brock symmetries with a Higgs like inducing of gravitational fields etc
which do not provide a complete and generally accepted reduction to the
general relativity theory. Our approach is in some lines ''inverse'' to the
former ones: we construct such nonholonomic distributions on a de Sitter
(principal) bundle $\ ^{N}\mathcal{S}a(\mathbf{V})=\left( \ ^{N}Sa(\mathbf{V}%
)\mathbf{,}Sf_{2+2+1},\mathbf{V}\right) ,$ for which the Einstein equations
in nonholonomic variables on a spacetime $\mathbf{V}$ induce equivalently
some nonholonomic gravitational gauge equations in the total space $\ ^{N}Sa(%
\mathbf{V}).$ This type of gravitational field equations are variational,
subjected to a class of nonholonomic constraints, and can be redefined,
after projection on a base spacetime into the system of field equations (\ref%
{einsteq}) (such base manifold equations are nonholonomic deformations of
the standard Einstein equations).

Let us consider a de Sitter space $\Sigma $ defined as a hypersurface $\eta
_{AB}u^{A}u^{B}=-1$ in a four--dimensional flat space endowed with a
diagonal metric $\eta _{AB}=diag[\pm 1,...,\pm 1],$ where $\{u^{A}\}$ are
global Cartezian coordinates in $\mathbb{R}^{5},$ indices $A,B,C...$ run
values $1,2,...,5$ and $l>0$ is the constant curvature of de Sitter space.
The isometry group of $\ ^{5}\Sigma $--space is the de Sitter group $\
^{\eta }S=\ SO(5).$ There are 6 operators $M_{AB}$ of Lie algebra $\mathit{\
so}(5)$ satisfying the commutation relations%
\begin{equation}
\lbrack M_{AB},M_{CD}]=\eta _{AC}M_{BD}-\eta _{BC}M_{AD}-\eta
_{AD}M_{BC}+\eta _{BD}M_{AC}.  \label{comrel}
\end{equation}%
A canonical $4+1$ splitting is parametrized by $A=(\underline{\alpha },5), B=(%
\underline{\beta },5),...; \eta _{AB}$ $=(\eta _{\underline{\alpha }\underline{%
\beta }}, \eta _{55})$ and $P_{\underline{\alpha }}=l^{-1}M_{5\underline{%
\alpha }},$ for $\underline{\alpha },\underline{\beta },...=1,2,3,4$ when the
commutation relations are written%
\begin{eqnarray}
\lbrack M_{\underline{\alpha }\underline{\beta }},M_{\underline{\gamma }%
\underline{\delta }}] &=&\eta _{\underline{\alpha }\underline{\gamma }}M_{%
\underline{\beta }\underline{\delta }}-\eta _{\underline{\beta }\underline{%
\gamma }}M_{\underline{\alpha }\underline{\delta }}+\eta _{\underline{\beta }%
\underline{\delta }}M_{\underline{\alpha }\underline{\gamma }}-\eta _{%
\underline{\alpha }\underline{\delta }}M_{\underline{\beta }\underline{%
\gamma }},  \label{lalgsr} \\
\lbrack P_{\underline{\alpha }},P_{\underline{\beta }}] &=&-l^{-2}M_{%
\underline{\alpha }\underline{\beta }},\ [P_{\underline{\alpha }},M_{%
\underline{\beta }\underline{\gamma }}]=\eta _{\underline{\alpha }\underline{%
\beta }}P_{\underline{\gamma }}-\eta _{\underline{\alpha }\underline{\gamma }%
}P_{\underline{\beta }}.  \notag
\end{eqnarray}%
The commutators (\ref{lalgsr}) are for a direct sum $\ \mathit{so}(5)=%
\mathit{so}(4)\oplus \ ^{4}V,$ where $\ ^{4}V$ is the four dimensional
vector space stretched on vectors $P_{\underline{\alpha }}.$ We remark that $%
\ ^{4}\Sigma =\ ^{\eta }S/\ ^{\eta }L,$ where $\ ^{\eta }L=\ SO(4).$
Choosing signature $\eta _{AC}=diag[-1,1,1,1,1]$ and $\ ^{\eta }S=SO(1,4),$
we get the group of Lorentz rotations $\ ^{\eta }L=SO(1,3).$ Here we also
note that the commutation relations (\ref{comrel}) define a non--Abelian Lie
algebra $\mathcal{A}_{I}$ \ for a gauge group with generators $I^{\underline{%
1}},...,I^{\underline{S}}$ and structure constants $f_{\underline{T}}^{\ \
\underline{S}\underline{P}}$ in $[I^{\underline{S}},I^{\underline{P}}]=if_{%
\underline{T}}^{\ \ \underline{S}\underline{P}}I^{\underline{T}}$
parametrized in a form to obtain the formulas (\ref{lalgsr}).\footnote{%
in order to preserve similarity with usual gauge field theories in particle
physics, it is convenient to use the complex unity ''$i$'' in Lie algebras
commutators even our nonholonomic gauge models are for the Einstein gravity
on real base spacetime manifolds}

At the next step, we consider $4\times 4$ matrix parametrizations of actions
of the group $\ ^{\eta }S$ adapted to splitting $\ \mathit{so}(5)=\mathit{so}%
(4)\oplus \ ^{4}V$ and distinguishing the subgroup $\ ^{\eta }L.$ We write
\begin{equation}
Q=q\ ^{L}Q,  \label{parametriz}
\end{equation}%
where$\ ^{L}Q=\left(
\begin{array}{cc}
L & 0 \\
0 & 1%
\end{array}%
\right) ,$ $q=\left(
\begin{array}{cc}
\delta _{\quad \underline{\beta }}^{\underline{\alpha }}+\frac{\tau ^{%
\underline{\alpha }}\tau _{\underline{\beta }}}{\left( 1+\tau ^{5}\right) }
& \tau ^{\underline{\alpha }} \\
\tau _{\underline{\beta }} & \tau ^{5}%
\end{array}%
\right)$ and $L\in \ ^{\eta }L$ is the de Sitter bust matrix transforming
the vector $\left( 0,0,...,\rho \right) \in \mathbb{R}^{5}$ into a point $%
\left( V^{1},V^{2},...,V^{5}\right) \in \ ^{5}\Sigma _{\rho }\subset \ ^{5}%
\mathcal{R},$ where $\ ^{5}\mathcal{R}$ is a 'sphere' with curvature $\rho ,$
$(V_{A}V^{A}=-\rho ^{2},V^{A}=\tau ^{A}\rho ).$

A de Sitter gauge field is given by a $\ \mathit{so}(5)$--valued connection
1--form
\begin{equation}
\widetilde{\Omega }=\left(
\begin{array}{cc}
\omega _{\quad \underline{\beta }}^{\underline{\alpha }} & \widetilde{\theta
}^{\underline{\alpha }} \\
\widetilde{\theta }_{\underline{\beta }} & 0%
\end{array}%
\right) ,  \label{dspot}
\end{equation}%
where $\omega _{\quad \underline{\beta }}^{\underline{\alpha }}\in \mathit{so%
}(4),$ $\widetilde{\theta }^{\underline{\alpha }}\in \mathcal{R}^{4},%
\widetilde{\theta }_{\underline{\beta }}\in \eta _{\underline{\beta }%
\underline{\alpha }}\widetilde{\theta }^{\underline{\alpha }}.$

The introduced para\-met\-ri\-za\-ti\-on is invariant on action of $SO\left(
4\right) $ group. So, it is not possible to say that $\omega _{\quad
\underline{\beta }}^{\underline{\alpha }}$ and $\widetilde{\theta }^{%
\underline{\alpha }}$ are induced directly by a d--connection $\mathbf{%
\Gamma }_{\ \beta \gamma }^{\alpha }=\ _{\shortmid }\Gamma _{\ \beta \gamma
}^{\alpha }-\ \mathbf{Z}_{\ \beta \gamma }^{\alpha }$ (\ref{condeform}) and
a 1--form $\mathbf{e}^{\alpha }$ (\ref{ddif}) because the actions of $\
^{\eta }S$ mix the components of the matrix $\omega _{\quad \beta ^{\prime
}}^{\alpha ^{\prime }}$ and $\widetilde{\theta }^{\alpha ^{\prime }}$ fields
in (\ref{dspot}).\footnote{%
that why we use primed indices and, following our conventions for spaces
with N--connections (in this section, de Sitter/affine frame bundles),
introduce boldface symbols} This problem can be solved on de Sitter affine \
bundles with N--connections and nonholonomic frames induced from the base
nonholonomic manifold. We also have to consider nonlinear gauge realizations
of the de Sitter group $\ ^{\eta }S$ when the nonlinear gauge field is
\begin{eqnarray}
\Gamma &=& q^{-1}{\widetilde{\Omega }}q+q^{-1}dq=\left(
\begin{array}{cc}
\mathbf{\Gamma }_{~\beta ^{\prime }}^{\alpha ^{\prime }} & \mathbf{\theta }%
^{\alpha ^{\prime }} \\
\mathbf{\theta }_{\beta ^{\prime }} & 0%
\end{array}%
\right) ,  \label{npot} \\
\mbox{ \ for \ }\mathbf{\Gamma }_{\quad \beta ^{\prime }}^{\alpha ^{\prime
}} &=&\omega _{\quad \beta ^{\prime }}^{\alpha ^{\prime }}-\left( \tau
^{\alpha ^{\prime }}\mathcal{D}\tau _{\beta ^{\prime }}-\tau _{\beta
^{\prime }}\mathcal{D}\tau ^{\alpha ^{\prime }}\right) /\left( 1+\tau
^{5}\right) ,  \notag \\
\theta ^{\alpha ^{\prime }} &=&\tau ^{5}\widetilde{\theta }^{\alpha ^{\prime
}}+\mathcal{D}\tau ^{\alpha ^{\prime }}-\tau ^{\alpha ^{\prime }}\left(
d\tau ^{5}+\widetilde{\theta }_{\gamma ^{\prime }}\tau ^{\gamma ^{\prime
}}\right) /\left( 1+\tau ^{5}\right) ,  \notag \\
\mathcal{D}\tau ^{\alpha ^{\prime }} &=&d\tau ^{\alpha ^{\prime }}+\omega
_{\quad \beta ^{\prime }}^{\alpha ^{\prime }}\tau ^{\beta ^{\prime }}.
\notag
\end{eqnarray}%
The action of the group $\ ^{\eta }S$ is nonlinear and nonholonomically
deformed,
\begin{equation*}
\Gamma ^{\prime }=L^{\prime }\Gamma \left( L^{\prime }\right)
^{-1}+L^{\prime }d\left( L^{\prime }\right) ^{-1},~\theta ^{\prime }=L\theta
,
\end{equation*}%
where the nonlinear matrix--valued function $L^{\prime }=L^{\prime }\left(
\tau ^{\alpha },q,Q\right) $ is defined from $Q_{q}=q^{\prime }\ ^{L^{\prime
}}Q$ (see the parametrization (\ref{parametriz})). The de Sitter 'nonlinear'
algebra is defined by generators (\ref{lalgsr}) and nonlinear gauge
transforms of type (\ref{npot}). In the language of nonholonomic
distributions, such nonlinear gauge transforms can be considered as certain
nonholonomic deformations of some linear gauge transforms.

In our approach, the de Sitter nonlinear gauge gravitational theory is to be
constructed from the coefficients of a d--metric $\mathbf{g}$ and
N--connection $\mathbf{N}$ in a form when the Einstein equations on the base
nonholonomic spacetime are equivalent to the Yang--Mills equations in the
total space. For this, we state that the d--connection (\ref{npot}) is
defined with respect to N--adapted frames (\ref{dder}) and (\ref{ddif}) and
a d--connection $\ ^{\mathbf{g}}\mathbf{\Gamma }_{\ \beta \gamma }^{\alpha
}, $
\begin{equation}
\ ^{\mathbf{g}}\mathbf{\Gamma }=\left(
\begin{array}{cc}
\ ^{\mathbf{g}}\mathbf{\Gamma }_{\quad \beta ^{\prime }}^{\alpha ^{\prime }}
& l_{0}^{-1}\mathbf{e}^{\alpha ^{\prime }} \\
l_{0}^{-1}\mathbf{e}_{\beta ^{\prime }} & 0%
\end{array}%
\right)  \label{conds}
\end{equation}%
where
\begin{equation}
\ ^{\mathbf{g}}\mathbf{\Gamma }_{\quad \beta ^{\prime }}^{\alpha ^{\prime
}}=\ ^{\mathbf{g}}\mathbf{\Gamma }_{\quad \beta ^{\prime }\mu }^{\alpha
^{\prime }}\delta u^{\mu },  \label{condsc}
\end{equation}%
for
\begin{equation}
\ ^{\mathbf{g}}\Gamma _{\quad \beta ^{\prime }\mu }^{\alpha ^{\prime }}=%
\mathbf{e}_{\alpha }^{~\alpha ^{\prime }}\mathbf{e}_{\quad \beta ^{\prime
}}^{\beta }\ ^{\mathbf{g}}\mathbf{\Gamma }_{\quad \beta \mu }^{\alpha }+%
\mathbf{e}_{\alpha }^{~\alpha ^{\prime }}\delta _{\mu }\mathbf{e}_{\quad
\beta ^{\prime }}^{\alpha },e^{\alpha ^{\prime }}=\mathbf{e}_{\mu }^{~\alpha
^{\prime }}\delta u^{\mu },  \label{coeff3}
\end{equation}%
with $l_{0}$ being a dimensional constant. The formulas (\ref{coeff3}) are
symilar to (\ref{dctr}) but, in this section, the indices $\alpha ^{\prime
},\beta ^{\prime }$ take values in the typical fiber/de Sitter space.

The matrix components of the curvature of the connection (\ref{conds}),
\begin{equation*}
\ ^{\Gamma }\mathcal{R}=d\ ^{\mathbf{g}}\mathbf{\Gamma }-\ ^{\mathbf{g}}%
\mathbf{\Gamma }\wedge \ ^{\mathbf{g}}\mathbf{\Gamma },
\end{equation*}%
can be parametrized in an invariant 4+1 form
\begin{equation}
\ ^{\Gamma }\mathcal{R}=\left(
\begin{array}{cc}
\mathcal{R}_{\quad \beta ^{\prime }}^{\alpha ^{\prime }}+l_{0}^{-1}\pi
_{\beta ^{\prime }}^{\alpha ^{\prime }} & l_{0}^{-1}\mathcal{T}^{\alpha
^{\prime }} \\
l_{0}^{-1}\mathcal{T}^{\beta ^{\prime }} & 0%
\end{array}%
\right) ,  \label{curvs}
\end{equation}
\begin{eqnarray*}
\mbox{ \ where \ }\pi _{\beta ^{\prime }}^{\alpha ^{\prime }} &=&\mathbf{e}%
^{\alpha ^{\prime }}\wedge \mathbf{e}_{\beta ^{\prime }},~\mathcal{T}^{\beta
^{\prime }}=\frac{1}{2}\ ^{\mathbf{g}}\mathbf{T}_{\quad \mu \nu }^{\beta
^{\prime }}\delta u^{\mu }\wedge \delta u^{\nu } \\
\mathcal{R}_{\quad \beta ^{\prime }}^{\alpha ^{\prime }} &=&\frac{1}{2}%
\mathcal{R}_{\quad \beta ^{\prime }\mu \nu }^{\alpha ^{\prime }}\delta
u^{\mu }\wedge \delta u^{\nu },\mathcal{R}_{\quad \beta ^{\prime }\mu \nu
}^{\alpha ^{\prime }}=\mathbf{e}_{~\beta ^{\prime }}^{\beta }\mathbf{e}%
_{\alpha }^{\quad \alpha ^{\prime }}\ ^{\mathbf{g}}\mathbf{R}_{\quad \beta
_{\mu \nu }}^{\alpha },\
\end{eqnarray*}%
when the coefficients $\ ^{\mathbf{g}}\mathbf{T}_{\quad \mu \nu }^{\beta
^{\prime }}$ are of type (\ref{dtors}) and $\ ^{\mathbf{g}}\mathbf{R}_{\quad
\beta {\mu \nu }}^{\alpha }$ are of type (\ref{dcurv}).

The de Sitter group is semisimple which allows us to to construct a
variational gauge gravitational theory with the Lagrangian
\begin{equation}
L=\ ^{\mathbf{g}}L+\ ^{m}L  \label{lagrangc}
\end{equation}%
where the gauge gravitational Lagrangian is
\begin{eqnarray*}
\ ^{\mathbf{g}}L &=&\frac{1}{4\pi }Tr\left( \ ^{\mathbf{\Gamma }}\mathcal{R}%
\wedge \ast \ ^{\mathbf{\Gamma }}\mathcal{R}\right) =\ ^{\mathbf{g}}\mathcal{%
L}\left| \mathbf{g}\right| ^{1/2}\delta ^{4}u, \\
\ ^{\mathbf{g}}\mathcal{L} &=&\frac{1}{2l^{2}}\ ^{\mathbf{g}}\mathbf{T}%
_{\quad \mu \nu }^{\alpha ^{\prime }}\ ^{\mathbf{g}}\mathbf{T}_{\alpha
^{\prime }}^{\quad \mu \nu }+\frac{1}{8\lambda }\mathcal{R}_{\quad \beta
^{\prime }\mu \nu }^{\alpha ^{\prime }}\mathcal{R}_{\quad \alpha ^{\prime
}}^{\beta ^{\prime }\quad \mu \nu }{}-\frac{1}{l^{2}}\left( {\overleftarrow{R%
}}-2\lambda _{1}\right) ,
\end{eqnarray*}%
for $\delta ^{4}u$ being the volume element. The Hodge operator '$\ast $' in
$\ ^{\mathbf{g}}L$ is constructed from the metric in the total space,
determinant $\left| \mathbf{g}\right| $ is computed from the coefficients of
a d--metric (\ref{block2}) stated with respect to N--elongated frames, the
curvature scalar ${\overleftarrow{R}}$ is computed as in (\ref{sdccurv}), $\
^{\mathbf{g}}\mathbf{T}_{\quad \mu \nu }^{\alpha ^{\prime }}=\mathbf{e}%
_{\quad \alpha }^{\alpha ^{\prime }}\ ^{\mathbf{g}}\mathbf{T}_{\quad \mu \nu
}^{\alpha }$ (the constant $l^{2}$ satisfies the relations $%
l^{2}=2l_{0}^{2}\lambda ,\lambda _{1}=-3/l_{0}),\quad Tr$ denotes the trace
on $\alpha ^{\prime },\beta ^{\prime }$ indices.

Let us discuss the meaning of constants (physical or non--physical ones)
introduced in the formulas of the above formulated nonholonomic gauge models
of gravity. We emphasize that such constants have different physical
interpretations in generalized gauge gravity theories and in the case of
nonholonomically gauge like structures induced by solutions of the Einstein
equations on the base spacetime. In the nonholonomic approach, the value $%
2\lambda _{1}/l^{2}$ can be considered as a usual cosmological constant $%
\Lambda /2$ in (\ref{seeq}). The constant $l^{2}$ characterizes the
intensity of torsion field interactions in a nonlinear de Sitter gauge
theory but it can be interpreted as a formal (nonphysical) constant defining
the nonholonomic constraints imposed on nonholonomic structure in order to
redefine the standard data $(\mathbf{g,}\ \ ^{\mathbf{g}}\nabla ),$ or $(%
\mathbf{g,\ }\ ^{\mathbf{g}}\mathbf{D}),$ for the Einstein gravity theory
into a nonholonomic de Sitter model with variables $\ ^{\mathbf{g}}\mathbf{%
\Gamma }$ (\ref{conds}). The constants $l^{2},l_{0}^{2}$ and $\lambda $ can
be included as particular cases of constant d--tensor $\mathbf{Y}$ fields
used in the Miron's procedure, see formula (\ref{mcdc}), generalized for
nonlinear de Sitter bundles. Such constants and fields characterize the type
of nonholonomic deformations of certain "usual Einstein data" into certain,
another type, but equivalent, "nonholonomic data". It is like we can chose
any equivalent, but convenient, frame or coordinate transform on a
spacetime, which for nonholonomic lifts on certain bundle spaces
characterize the properties of such lifts keeping the proprieties of
"Einstein data" even they are nonholonomically deformed under such lifts. In
a more general context, such constructions reflect the fact that
nonholonomic distributions provide more rich geometric and physical
structures, and a number of more sophisticate geometric methods and
constructions, than in the case of unconstrained dynamics of gravitational
fields. Such constructions provide more possibilities in developing quantum
models, but also for applications of classical gravity in cosmology and
astrophysics because of a multi--connection character of theories with
nonholonomic distributions\footnote{%
all connections under consideration being defined by the same metric but
with possible additional interaction constants, nonholonomic constraints and
prescribed symmetries and paramatetrizations; from formal point of view, we
can redefine the geometric constructions for the Levi--Civita connection but
this results in a more sophisticate "mixture" of nonholonomic deformations
and constraints, nonlinear interactions and broken symmetries and associated
constants and/or chosen parametrizations of fields}.

The matter field Lagrangian from (\ref{lagrangc}) is
\begin{equation*}
\ ^{m}L=-\frac{1}{2}Tr\left( \ ^{\mathbf{g}}\mathbf{\Gamma }\wedge \ast _{%
\mathbf{g}}\mathcal{I}\right) =\ ^{m}\mathcal{L}\left| \mathbf{g}\right|
^{1/2}\delta ^{4}u,
\end{equation*}%
when the Hodge operator $\ast _{\mathbf{g}}$ is defined by $\left| \mathbf{g}%
\right| ,$ where
\begin{equation*}
\ ^{m}\mathcal{L}=\frac{1}{2}\ ^{\mathbf{g}}\mathbf{\Gamma }_{\quad \beta
^{\prime }\mu }^{\alpha ^{\prime }}\mathcal{S}_{\quad \alpha }^{\beta
^{\prime }\quad \mu }-\tau _{\quad \alpha ^{\prime }}^{\mu }\mathbf{e}%
_{\quad \mu }^{\alpha ^{\prime }}.
\end{equation*}%
The matter field source $\mathcal{J}$ is constructed as a variational
derivation of $\ ^{m}\mathcal{L}$ on $\ ^{\mathbf{g}}\Gamma $ and is
paramet\-riz\-ed in the form
\begin{equation*}
\mathcal{J}=\left(
\begin{array}{cc}
\mathcal{S}_{\quad \underline{\beta }}^{\alpha ^{\prime }} & -l_{0}\tau
^{\alpha ^{\prime }} \\
-l_{0}\tau _{\beta ^{\prime }} & 0%
\end{array}%
\right) ,
\end{equation*}%
with $\tau ^{\alpha ^{\prime }}=\tau _{\quad \mu }^{\alpha ^{\prime }}\delta
u^{\mu }$ and $\mathcal{S}_{\quad \beta ^{\prime }}^{\alpha ^{\prime }}=%
\mathcal{S}_{\quad \beta ^{\prime }\mu }^{\alpha ^{\prime }}\delta u^{\mu }$
being respectively the canonical tensors of energy--momentum and spin
density.

Varying the action $\ S=\int \delta ^{4}u\left( \ ^{\mathbf{g}}\mathcal{L}+\
^{m}\mathcal{L}\right) $ on the $\Gamma $--variables (\ref{conds}), we
obtain the gau\-ge--gra\-vi\-ta\-ti\-on\-al field equations:%
\begin{equation}
d\left( \ast \ ^{\mathbf{\Gamma }}\mathcal{R}\right) +\ ^{\mathbf{g}}\Gamma
\wedge \left( \ast \ ^{\mathbf{\Gamma }}\mathcal{R}\right) -\left( \ast \ ^{%
\mathbf{\Gamma }}\mathcal{R}\right) \wedge \ ^{\mathbf{g}}\Gamma =-\lambda
\left( \ast \mathcal{J}\right) .  \label{eqs}
\end{equation}%
This equations can be alternatively derived in geometric form by applying
the absolute derivation and dual operators.

Distinguishing the variations on $\ ^{\mathbf{g}}\mathbf{\Gamma }$ and $%
\mathbf{e}$--variables, we rewrite (\ref{eqs})
\begin{eqnarray}
\widetilde{\mathcal{D}}\left( \ast \ ^{\mathbf{\Gamma }}\mathcal{R}\right) +%
\frac{2\lambda }{l^{2}}\left( \widetilde{\mathcal{D}}\left( \ast \pi \right)
+\mathbf{e}\wedge \ast \mathbf{T}^{t}-\ast \mathbf{T}\wedge \mathbf{e}%
^{t}\right) &=&-\lambda \left( \ast \mathcal{S}\right) ,  \label{heq1} \\
\widetilde{\mathcal{D}}\left( \ast \mathbf{T}\right) -\left( \ast \ ^{%
\mathbf{\Gamma }}\mathcal{R}\right) \wedge e-\frac{2\lambda }{l^{2}}\ast \pi
\wedge \mathbf{e} &=&\frac{l^{2}}{2}\left( \ast \tau +\frac{1}{\lambda }\ast
\varsigma \right) ,  \notag
\end{eqnarray}%
$\mathbf{e}^{t}$ being the transposition of \ $\mathbf{e},$ where
\begin{equation}
\mathbf{T}^{t}=\{~\mathcal{T}_{\alpha ^{\prime }}=\eta _{\alpha ^{\prime
}\beta ^{\prime }}~\mathcal{T}^{\beta ^{\prime }}\},\ \mathbf{e}^{T}=\{%
\mathbf{e}_{\alpha ^{\prime }}=\eta _{\alpha ^{\prime }\beta ^{\prime }}%
\mathbf{e}^{\beta ^{\prime }},~\mathbf{e}^{\beta ^{\prime }}=\mathbf{e}%
_{\quad \mu }^{\beta ^{\prime }}\delta u^{\mu }\},  \notag
\end{equation}%
for $\widetilde{\mathcal{D}}=\delta +\widetilde{\Gamma };$ the operators $%
\widetilde{\Gamma }$ acts as $\ ^{\mathbf{g}}\mathbf{\Gamma }_{\quad \beta
^{\prime }\mu }^{\alpha ^{\prime }}$ on indices $\gamma ^{\prime },\delta
^{\prime },...$ and as $\ ^{\mathbf{g}}\mathbf{\Gamma }_{\quad \beta \mu
}^{\alpha }$ on indices $\gamma ,\delta ,....$ For this model, we can
construct an energy--momentum tensor for the gauge gravitational field $%
\widetilde{\Gamma },$%
\begin{equation}
\varsigma _{\mu \nu }\left( \widetilde{\Gamma }\right) =\frac{1}{2}Tr\left(
\ ^{\mathbf{\Gamma }}\mathcal{R}_{\mu \alpha }\ ^{\mathbf{\Gamma }}\mathcal{R%
}_{\quad \nu }^{\alpha }-\frac{1}{4}\ ^{\mathbf{\Gamma }}\mathcal{R}_{\alpha
\beta }\ ^{\mathbf{\Gamma }}\mathcal{R}^{\alpha \beta }\mathbf{g}_{\mu \nu
}\right) .  \label{gemgf}
\end{equation}

Equations (\ref{eqs}) make up the complete system of variational field
equations for nonlinear de Sitter gauge gravity. We note that we can obtain
a nonvariational Poincar\' e gauge gravitational theory if we consider the
contraction of the gauge potential (\ref{conds}) to the 1--form (\ref{aymf}%
), i.e. $\ \ $
\begin{equation*}
\ ^{\mathbf{g}}\mathbf{\Gamma }=\left(
\begin{array}{cc}
\ ^{\mathbf{g}}\mathbf{\Gamma }_{\quad \beta ^{\prime }}^{\alpha ^{\prime }}
& l_{0}^{-1}\mathbf{e}^{\alpha ^{\prime }} \\
l_{0}^{-1}\mathbf{e}_{\beta ^{\prime }} & 0%
\end{array}%
\right) \rightarrow \overline{\mathbf{\omega }}=\ ^{\mathbf{g}}\overline{%
\Gamma }.
\end{equation*}%
For $\ ^{\mathbf{g}}\mathbf{\Gamma ,}$ or $\ ^{\mathbf{g}}\overline{\Gamma }%
, $ projected on the base nonholonomic spacetime, our nonholonomic de Sitter
gauge equations are completely equivalent to the Einstein equations (\ref%
{einsteq}), and (reintroducing the Levi--Civita connection) to (\ref{seeq}).
We have to use deformations of type $\ \mathbf{\Gamma }_{\ \beta \gamma
}^{\alpha }=\ _{\shortmid }\Gamma _{\ \beta \gamma }^{\alpha }-\ \mathbf{Z}%
_{\ \beta \gamma }^{\alpha }$ and $\mathcal{R}_{\ \gamma }^{\tau }=\
_{\shortmid }\mathcal{R}_{\ \gamma }^{\tau }-\mathcal{Z}_{\ \gamma }^{\tau
}, $ where $\ _{\shortmid }\mathcal{R}_{\ \gamma }^{\tau }$ $=\ _{\shortmid
}R_{\ \gamma \alpha \beta }^{\tau }\ \mathbf{e}^{\alpha }\wedge \ \mathbf{e}%
^{\beta }$ is the curvature 2--form of the Levi--Civita connection $\nabla $
and $\mathcal{R}_{\ \gamma }^{\tau }=\mathbf{R}_{\ \gamma \alpha \beta
}^{\tau }\ \mathbf{e}^{\alpha }\wedge \ \mathbf{e}^{\beta }$ $\ $the
curvature tensor of a $\mathbf{D,}$ related distorsion of curvature 2--form $%
\mathcal{Z}_{\ \gamma }^{\tau }$ defined by $\ \mathbf{Z}_{\ \beta \gamma
}^{\alpha }$ (\ref{distgf}). We have to include such terms in an induced
spin source $\mathcal{S}_{\quad \beta ^{\prime }\mu }^{\alpha ^{\prime }},$
similarly as we induced the distorsion $\ ^{Z}T_{\quad \mu }^{\alpha
^{\prime }}$ of matter energy--momentum tensor for (\ref{ident}).

It is a well known fact that in general relativity it is not possible to
define a usual energy--momentum tensor for the gravitational field.
Nevertheless, for its nonholonomic deformations induced on some bundle
spaces such tensors can be defined at least formally, see (\ref{gemgf}). For
nonholonomic distributions with $3+1$ and $2+2$ spacetime decompositions
(for a de Sitter extension, we have to consider a $3+2$ splitting) certain
constructions with effective gravitational energy and momenta, like in \cite%
{asht,rov,thiem1}, for usual $3+1$ variables in classical and quantum
gravity, can be performed.

\subsection{Gravitational Yang--Mills equations for distorsion d--ten\-sors}

The d--connection and d--tensor formalism allows us to adapt the geometric
constructions to a prescribed nonholonomic distribution (N--connection)
structure. N--anholonomic distributions can be defined in arbitrary form on
a spacetime manifold and bundle spaces on such a manifold like we are free
to chose any frame or coordinate structure on a curved spacetime. But we can
also chose a nonholonomic distribution, with corresponding constraints on
dynamics of nonlinearly interacting gravitational and matter fields, when
certain configurations of fields and nonintegrable constraints became stable
or subjected to generate certain interactions and evolutions of fields and
geometries with prescribed symmetry and/or nonholonomic deformations \cite%
{vnhrf1,vnhrf2}.

The fact that from a given metric tensor on a nonholonomic manifold one can
be constructed different classes of linear connections, all defined by the
same metric structure, allows us to develop us new models of gauge
gravitational interactions which are equivalents of the Einstein gravity
theory but with a more rich nonholonomic geometric structure.

Let us write the gauge like equivalent of Einstein equations (\ref{pdeq}),
in the bundle of affine frames $\ \mathcal{A}a(\mathbf{V})=\left( \ Aa(%
\mathbf{V})\mathbf{,}Af_{4},\mathbf{V}\right) $ on a (pseudo) Riemannian
spacetime $\mathbf{V},$ in the form\
\begin{equation}
d\left( \ast \ \ _{\shortmid }^{\mathbf{\omega }}\overline{\mathcal{R}}%
\right) +\ _{\shortmid }^{\mathbf{g}}\overline{\Gamma }\wedge \left( \ast \
\ _{\shortmid }^{\mathbf{\omega }}\overline{\mathcal{R}}\right) -\left( \ast
\ \ _{\shortmid }^{\mathbf{\omega }}\overline{\mathcal{R}}\right) \wedge \ ^{%
\mathbf{g}}\overline{\Gamma }=-\ _{\shortmid }^{\mathbf{g}}\overline{%
\mathcal{J}}.  \label{pd1a}
\end{equation}%
for a gravitational connection $\overline{\mathbf{\omega }}=\ _{\shortmid }^{%
\mathbf{g}}\overline{\Gamma }$ (\ref{afgym}) defined by the Levi--Civita
connection and $\ _{\shortmid }^{\mathbf{g}}\overline{\mathcal{J}}$ induced
by the energy--momentum tensor of matter fields in general relativity.%
\footnote{%
we emphasize that similar constructions, but for different nonholonomic de
Sitter distributions, can be performed if we begin with the gauge gravity
equations (\ref{eqs}); for simplicity, in this section, we shall consider
only nonholonomic affine frame bundles and connections on such spaces} Any
nonholonomic splitting on $\mathbf{V}$ of type $\ _{\shortmid }\Gamma _{\
\beta \gamma }^{\alpha }=$ $\mathbf{\Gamma }_{\ \beta \gamma }^{\alpha }+\
_{\shortmid }\mathbf{Z}_{\ \beta \gamma }^{\alpha },$ see formulas (\ref%
{condeform}) and (\ref{distgf}), results in \ $\ _{\shortmid }\mathcal{R}_{\
\gamma }^{\tau }$ $=\mathcal{R}_{\ \gamma }^{\tau }+\ _{\shortmid }\mathcal{Z%
}_{\ \gamma }^{\tau },$ i.e. $\ _{\shortmid }^{\mathbf{g}}\overline{\Gamma }%
=\ ^{\mathbf{g}}\overline{\mathbf{\Gamma }}+\ _{\shortmid }^{\mathbf{g}}%
\overline{\mathbf{Z}}$ \ generates $\ \ _{\shortmid }^{\mathbf{\omega }}%
\overline{\mathcal{R}}=\ ^{\mathbf{g}}\overline{\mathcal{R}}+\ _{\shortmid
}^{\mathbf{g}}\overline{\mathcal{Z}}.$ Introducing such distorsion relations
into (\ref{pd1a}), for a curvature $\ ^{\mathbf{g}}\overline{\mathcal{R}}$
solving the equations
\begin{equation*}
d\left( \ast \ ^{\mathbf{g}}\overline{\mathcal{R}}\right) +\ ^{\mathbf{g}}%
\overline{\Gamma }\wedge \left( \ast \ ^{\mathbf{g}}\overline{\mathcal{R}}%
\right) -\left( \ast \ ^{\mathbf{g}}\overline{\mathcal{R}}\right) \wedge \ ^{%
\mathbf{g}}\overline{\Gamma }=0,
\end{equation*}%
(they are equivalent to equations (\ref{deinsteq}) for $\Lambda =0$ and $%
\overleftrightarrow{\mathbf{T}}_{\ \beta }^{\underline{\alpha }}=0$), we get%
\begin{equation}
d\left( \ast \ \ _{\shortmid }^{\mathbf{g}}\overline{\mathcal{Z}}\right) +\
\ _{\shortmid }^{\mathbf{g}}\overline{\mathbf{Z}}\wedge \left( \ast \ \
_{\shortmid }^{\mathbf{g}}\overline{\mathcal{Z}}\right) -\left( \ast \ \ \
_{\shortmid }^{\mathbf{g}}\overline{\mathcal{Z}}\right) \wedge \ \
_{\shortmid }^{\mathbf{g}}\overline{\mathbf{Z}}=-\ _{\shortmid }^{\mathbf{z}}%
\overline{\mathcal{J}},  \label{efymeq}
\end{equation}%
where the nonholonomically deformed source is%
\begin{equation*}
\ _{\shortmid }^{\mathbf{z}}\overline{\mathcal{J}}=\ _{\shortmid }^{\mathbf{g%
}}\overline{\mathcal{J}}+\ ^{\mathbf{g}}\overline{\Gamma }\wedge \left( \ast
\ ^{\mathbf{g}}\overline{\mathcal{Z}}\right) -\left( \ast \ ^{\mathbf{g}}%
\overline{\mathcal{Z}}\right) \wedge \ ^{\mathbf{g}}\overline{\Gamma }+\ \
_{\shortmid }^{\mathbf{g}}\overline{\mathbf{Z}}\wedge \left( \ast \ ^{%
\mathbf{g}}\overline{\mathcal{R}}\right) -\left( \ast \ ^{\mathbf{g}}%
\overline{\mathcal{R}}\right) \wedge \ \ _{\shortmid }^{\mathbf{g}}\overline{%
\mathbf{Z}}.
\end{equation*}%
Source $\ _{\shortmid }^{\mathbf{z}}\overline{\mathcal{J}}$ is the total
space analog of base source $\mathcal{T}_{\ \tau }=\ ^{m}\mathcal{T}_{\ \tau
}+\ ^{Z}\mathcal{T}_{\ \tau }$ for the Einstein equations in nonholonomic
variables (\ref{einsteq}). The induced from base N--anholonomic distribution
in $\ Aa(\mathbf{V})$ states a nonlinear relation between the distorsion of
curvature $\ \ _{\shortmid }^{\mathbf{g}}\overline{\mathcal{Z}} $ and
distorsion of the Cartan d--connection $\ _{\shortmid }^{\mathbf{g}}%
\overline{\mathbf{Z}},$ which is different from that in standard Yang--Mills
theory, of type (\ref{curvafb}).

The effective Yang--Mills gravitational equations and their source can be
substantially simplified for nonholonomic distributions when $\mathbf{g=%
\mathring{g}}$ on base spacetime $\mathbf{V}$ induces a d--connection $%
\widehat{\mathbf{\mathring{\Gamma}}}_{\ \alpha ^{\prime }\beta ^{\prime
}}^{\gamma ^{\prime }}=(0,\widehat{\mathring{L}}_{\ b^{\prime }k^{\prime
}}^{a^{\prime }},0,0)$ (\ref{ccandcon}) with constant curvature coefficients
$\ \widehat{\mathbf{\mathring{R}}}_{\ \beta ^{\prime }\gamma ^{\prime
}\delta ^{\prime }}^{\alpha ^{\prime }}=(0,\widehat{\mathring{R}}%
_{~b^{\prime }j^{\prime }k^{\prime }}^{a^{\prime }}=\ \widehat{\mathring{L}}%
_{\ b^{\prime }j^{\prime }}^{c^{\prime }}\ \widehat{\mathring{L}}_{\
c^{\prime }k^{\prime }}^{a^{\prime }}-\ \widehat{\mathring{L}}_{\ b^{\prime
}k^{\prime }}^{c^{\prime }}\ \widehat{\mathring{L}}_{\ c^{\prime }j^{\prime
}}^{a^{\prime }},0,0,0,0)$ (\ref{ccdcc}), with respect to a class of
N--adapted frames. The corresponding distorsion of the Levi--Civita
connection with respect to $\widehat{\mathbf{\mathring{\Gamma}}}_{\ \alpha
^{\prime }\beta ^{\prime }}^{\gamma ^{\prime }}$ is written in the form $\
_{\shortmid }\Gamma _{\ \beta \gamma }^{\alpha }=\widehat{\mathbf{\mathring{%
\Gamma}}}_{\ \beta \gamma }^{\alpha }+\ _{\shortmid }\widehat{\mathbf{%
\mathring{Z}}}_{\ \beta \gamma }^{\alpha }.$ The related distorsions in the
total space are $\ _{\shortmid }^{\mathbf{g}}\overline{\Gamma }=\ \overline{%
\mathbf{\mathring{\Gamma}}}+\ _{\shortmid }\ \overline{\mathbf{\mathring{Z}}}
$ \ and $\ \ _{\shortmid }^{\mathbf{\omega }}\overline{\mathcal{R}}=\
\overline{\mathcal{\mathring{R}}}+\ _{\shortmid }\overline{\mathcal{%
\mathring{Z}}}.$

For a four dimensional nonholonomic (pseudo) Riemannian base $\mathbf{V,}$
one could be maximum eight nontrivial components $\widehat{\mathring{L}}_{\
b^{\prime }k^{\prime }}^{a^{\prime }}.$ We can prescribe such a nonholonomic
distribution with some nontrivial values $\widehat{\mathring{L}}_{\
b^{\prime }j^{\prime }}^{c^{\prime }}$ when $\widehat{\mathring{R}}%
_{~b^{\prime }j^{\prime }k^{\prime }}^{a^{\prime }}=\ \widehat{\mathring{L}}%
_{\ b^{\prime }j^{\prime }}^{c^{\prime }}\ \widehat{\mathring{L}}_{\
c^{\prime }k^{\prime }}^{a^{\prime }}-\ \widehat{\mathring{L}}_{\ b^{\prime
}k^{\prime }}^{c^{\prime }}\ \widehat{\mathring{L}}_{\ c^{\prime }j^{\prime
}}^{a^{\prime }}=0.$\footnote{%
It should be emphasized here that the geometric properties of curvature
d--tensor for a d--connection are very different from those of usual tensors
and linear connections. Even the coefficients of a d--tensor may vanish with
respect to a particular class of nonholonomic distributions, the real
spacetime may be a general (pseudo) Riemannian one with nontrivial curvature
of the Levi--Civita connection and nonzero associated/induced
nonholonomically d--torsions, nonholonomy coefficients and curvature of
N--connection.} Choosing $\ \overline{\mathcal{\mathring{R}}}=0,$ we write
the gauge like gravitational equations (\ref{efymeq}) in a simplified form,%
\begin{equation}
d\left( \ast \ _{\shortmid }\overline{\mathcal{\mathring{Z}}}\right) +\
_{\shortmid }\ \overline{\mathbf{\mathring{Z}}}\wedge \left( \ast \ \
_{\shortmid }\overline{\mathcal{\mathring{Z}}}\right) -\left( \ast \ \
_{\shortmid }\overline{\mathcal{\mathring{Z}}}\right) \wedge \ \ _{\shortmid
}\ \overline{\mathbf{\mathring{Z}}}=-\ _{\shortmid }\overline{\mathcal{%
\mathring{J}}},  \label{efymeq1}
\end{equation}%
where the nonholonomically deformed source is%
\begin{equation*}
\ _{\shortmid }\overline{\mathcal{\mathring{J}}}=\ _{\shortmid }^{\mathbf{g}}%
\overline{\mathcal{J}}+\ \overline{\mathbf{\mathring{\Gamma}}}\wedge \left(
\ast \ _{\shortmid }\overline{\mathcal{\mathring{Z}}}\right) -\left( \ast \
_{\shortmid }\overline{\mathcal{\mathring{Z}}}\right) \wedge \ \overline{%
\mathbf{\mathring{\Gamma}}}.
\end{equation*}%
Induced gauge like gravitational equations of type (\ref{efymeq}), or (\ref%
{efymeq1}), with constraints of type $\ \widehat{\mathbf{\mathring{R}}}_{\
\beta ^{\prime }\gamma ^{\prime }\delta ^{\prime }}^{\alpha ^{\prime
}}=const,$ or $=0,$ are important for elaborating a formal scheme for
perturbative quantization and renormalization of general relativity using
the so--called ''nonholonomic two--connection formalism'' in \cite{vpqgg}.

\section{Concluding Remarks}

In this work the Einstein gravity theory was reformulated in some different,
but equivalent, forms on manifolds and affine/de Sitter frame bundles
enabled with nonholonomic distributions. There were considered distributions
with 2+2 splitting of base spacetime and associated nonholonomic frame
structures. Using the formalism of nonlinear connections, the geometric
constructions were adapted to formal decompositions into holonomic and
nonholonomic variables.

We followed two guiding geometric principles: 1) for a (pseudo) Riemannian
metric, it is possible to construct an infinite number of metric compatible
linear connections, all determined by a metric structure in unique forms
following certain well defined geometric/physical conditions; 2) imposing
certain types of nonintegrable/nonholonomic constraints, we can select some
classes of connections and nonholonomic deformations of geometric objects
which are most convenient for various purposes in modern gravity (for
instance, to define some invariants and conservation laws, to elaborate new
quantization schemes for gravity theory etc); all such geometric
constructions can be equivalently re--defined in terms of a Levi--Civita
connection.

The gauge gravitational models presented here are related to a number of
attempts to find a Yang--Mills formulation/ generalization for Einstein
gravity. The bulk of former constructions were oriented to elaborate various
gauge gravity theories which in certain limits model classical and possible
quantum effects for general relativity. In our approach, we considered some
inverse constructions. We defined such nonholonomic distibutions on a base
spacetime, and their lifts on bundle spaces, when the geometric data for a
solution of Einstein equations can be encoded equivalently into some classes
of nonholonomic frames and linear and nonlinear connections stating the
geometric structure of such bundles.

The formulated equivalents of the Einstein equations on nonholonomic
manifolds and (affine/de Sitter) frame bundle spaces contain non--trivial
torsion contributions. For various generalizations of Einstein gravity to
other gauge/ string / Riemann--Cartan gravity theories, the torsion fields
are different from the metric one and subjected to certain additional
field/algebraic equations. In our approach, the torsion structure is induced
nonholonomically by certain off--diagonal metric (equivalently, nonholonomy)
coefficients. So, we do not generalize the Einstein gravity theory, but
represent it equivalently in terms of some new types of nonholonomic
variables, which is more convenient for certain constructions in quantum
gravity (preserving an analogy with the usual Yang--Mills fields) and, for
instance, to develop new geometric methods for constructing exact solutions
in gravity and Ricci flow theories (see new results and reviews in our works %
\cite{ijgmmp,vrflg,vncg,vacap,vpla,anav,vloopdq,vgwgr,vnhrf1}).

Finally, we note that this paper provides a nonholonomic geometric formalism
for a second partner works on a model of perturbative quantum gauge gravity \cite%
{vpqgg}.

\vskip5pt

\textbf{Acknowledgement: } The work was partially performed
during a visit at the Fields Institute.

\setcounter{equation}{0} \renewcommand{\theequation}
{A.\arabic{equation}} \setcounter{subsection}{0}
\renewcommand{\thesubsection}
{A.\arabic{subsection}}

\appendix

\section{Coefficients of d--Connections and d--Tensors}

For convenience, in this Appendix, we outline some important coefficient
formulas for fundamental geometric objects on N--anholonomic manifolds.
Details and proofs are presented in Refs. \cite{vrflg,vncg,vacap}.

Locally it is characterized by (N--adapted) d--torsion coefficients
\begin{eqnarray}
T_{\ jk}^{i} &=&L_{\ jk}^{i}-L_{\ kj}^{i},\ T_{\ ja}^{i}=-T_{\ aj}^{i}=C_{\
ja}^{i},\ T_{\ ji}^{a}=\Omega _{\ ji}^{a},\   \label{dtors} \\
T_{\ bi}^{a} &=&-T_{\ ib}^{a}=\frac{\partial N_{i}^{a}}{\partial y^{b}}-L_{\
bi}^{a},\ T_{\ bc}^{a}=C_{\ bc}^{a}-C_{\ cb}^{a}.  \notag
\end{eqnarray}

The curvature of a d--connection $\mathbf{D,}$
\begin{equation}
\mathcal{R}_{~\beta }^{\alpha }\doteqdot \mathbf{D\Gamma }_{\ \beta
}^{\alpha }=d\mathbf{\Gamma }_{\ \beta }^{\alpha }-\mathbf{\Gamma }_{\ \beta
}^{\gamma }\wedge \mathbf{\Gamma }_{\ \gamma }^{\alpha },  \label{curv}
\end{equation}%
splits into six types of N--adapted components with respect to (\ref{dder})
and (\ref{ddif}),
\begin{equation*}
\mathbf{R}_{~\beta \gamma \delta }^{\alpha }=\left(
R_{~hjk}^{i},R_{~bjk}^{a},P_{~hja}^{i},P_{~bja}^{c},S_{~jbc}^{i},S_{~bdc}^{a}\right) ,
\end{equation*}%
\begin{eqnarray}
R_{\ hjk}^{i} &=&\mathbf{e}_{k}L_{\ hj}^{i}-\mathbf{e}_{j}L_{\ hk}^{i}+L_{\
hj}^{m}L_{\ mk}^{i}-L_{\ hk}^{m}L_{\ mj}^{i}-C_{\ ha}^{i}\Omega _{\ kj}^{a},
\label{dcurv} \\
R_{\ bjk}^{a} &=&\mathbf{e}_{k}L_{\ bj}^{a}-\mathbf{e}_{j}L_{\ bk}^{a}+L_{\
bj}^{c}L_{\ ck}^{a}-L_{\ bk}^{c}L_{\ cj}^{a}-C_{\ bc}^{a}\Omega _{\ kj}^{c},
\notag \\
P_{\ jka}^{i} &=&e_{a}L_{\ jk}^{i}-D_{k}C_{\ ja}^{i}+C_{\ jb}^{i}T_{\
ka}^{b},  \notag \\
P_{\ bka}^{c} &=&e_{a}L_{\ bk}^{c}-D_{k}C_{\ ba}^{c}+C_{\ bd}^{c}T_{\
ka}^{c},  \notag \\
S_{\ jbc}^{i} &=&e_{c}C_{\ jb}^{i}-e_{b}C_{\ jc}^{i}+C_{\ jb}^{h}C_{\
hc}^{i}-C_{\ jc}^{h}C_{\ hb}^{i},  \notag \\
S_{\ bcd}^{a} &=&e_{d}C_{\ bc}^{a}-e_{c}C_{\ bd}^{a}+C_{\ bc}^{e}C_{\
ed}^{a}-C_{\ bd}^{e}C_{\ ec}^{a}.  \notag
\end{eqnarray}

Contracting respectively the components, $\mathbf{R}_{\alpha \beta
}\doteqdot \mathbf{R}_{\ \alpha \beta \tau }^{\tau },$ one computes the h-
v--components of the Ricci d--tensor
\begin{equation}
R_{ij}\doteqdot R_{\ ijk}^{k},\ \ R_{ia}\doteqdot -P_{\ ika}^{k},\
R_{ai}\doteqdot P_{\ aib}^{b},\ S_{ab}\doteqdot S_{\ abc}^{c}.
\label{dricci}
\end{equation}%
The scalar curvature is defined by contracting the Ricci d--tensor with the
inverse metric $\mathbf{g}^{\alpha \beta },$
\begin{equation}
\overleftrightarrow{\mathbf{R}}\doteqdot \mathbf{g}^{\alpha \beta }\mathbf{R}%
_{\alpha \beta }=g^{ij}R_{ij}+h^{ab}S_{ab}=\overrightarrow{R}+\overleftarrow{%
S}.  \label{sdccurv}
\end{equation}

Finally, we emphasize that formulas presented in this section hold true both
for any d--connection and/or not N--adapted linear connection (in the last
case, one should be not used 'boldface' symbols) with respect to arbitrary
(non) holonomic frames. Working with metric compatible linear connections,
we get coefficient formulas which are very similar to those for the standard
(in (pseudo) Riemannian geometry) Levi--Civita connection. Nevertheless,
nonholonomic configurations may change certain symmetry properties of
geometric objects and physical equations/conservation laws. For instance,
the Ricci d--tensor (\ref{dricci}), in general, is not symmetric, i.e. $%
\mathbf{R}_{\alpha \beta }\neq \mathbf{R}_{\beta \alpha }$ even for
symmetric metrics. Working with d--connections/linear connections completely
defined by a metric structure, we can always redefine the constructions in
an equivalent 'Levi--Civita form' when the usual symmetries and Bianchi
identities are present.

Any geometric construction for the canonical d--connection $\widehat{\mathbf{%
D}}$ (\ref{candcon}) can be re--defined for the Levi--Civita connection $%
\bigtriangledown =\{\ _{\shortmid }\Gamma _{\beta \gamma }^{\alpha }\}$ by
using the formula
\begin{equation}
\ _{\shortmid }\Gamma _{\ \alpha \beta }^{\gamma }=\widehat{\mathbf{\Gamma }}%
_{\ \alpha \beta }^{\gamma }+\ _{\shortmid }Z_{\ \alpha \beta }^{\gamma },
\label{deflc}
\end{equation}%
where the both connections $\ _{\shortmid }\Gamma _{\ \alpha \beta }^{\gamma
}$ and $\widehat{\mathbf{\Gamma }}_{\ \alpha \beta }^{\gamma }$ and the
distorsion tensor $\ _{\shortmid }Z_{\ \alpha \beta }^{\gamma }$ with
N--adapted coefficients where%
\begin{eqnarray}
\ _{\shortmid }Z_{jk}^{a} &=&-\widehat{C}_{jb}^{i}g_{ik}h^{ab}-\frac{1}{2}%
\Omega _{jk}^{a},~_{\shortmid }Z_{bk}^{i}=\frac{1}{2}\Omega
_{jk}^{c}h_{cb}g^{ji}-\Xi _{jk}^{ih}~\widehat{C}_{hb}^{j},  \notag \\
_{\shortmid }Z_{bk}^{a} &=&~^{+}\Xi _{cd}^{ab}~~^{\bullet }\widehat{L}%
_{bk}^{c},\ _{\shortmid }Z_{kb}^{i}=\frac{1}{2}\Omega
_{jk}^{a}h_{cb}g^{ji}+\Xi _{jk}^{ih}~\widehat{C}_{hb}^{j},\ _{\shortmid
}Z_{jk}^{i}=0,  \label{deft} \\
\ _{\shortmid }Z_{jb}^{a} &=&-~^{-}\Xi _{cb}^{ad}~~^{\bullet }\widehat{L}%
_{dj}^{c},\ _{\shortmid }Z_{bc}^{a}=0,\ _{\shortmid }Z_{ab}^{i}=-\frac{g^{ij}%
}{2}\left[ ~^{\bullet }\widehat{L}_{aj}^{c}h_{cb}+~^{\bullet }\widehat{L}%
_{bj}^{c}h_{ca}\right] ,  \notag
\end{eqnarray}%
for $\ \Xi _{jk}^{ih}=\frac{1}{2}(\delta _{j}^{i}\delta
_{k}^{h}-g_{jk}g^{ih}),~^{\pm }\Xi _{cd}^{ab}=\frac{1}{2}(\delta
_{c}^{a}\delta _{d}^{b}+h_{cd}h^{ab})$ and$~^{\bullet }\widehat{L}_{aj}^{c}=%
\widehat{L}_{aj}^{c}-e_{a}(N_{j}^{c}).$ If we work with nonholonomic
constraints on the dynamics/ geometry of gravity fields, it is more
convenient to use a N--adapted approach. For other purposes, it is preferred
to use only the Levi--Civita connection.

\end{document}